\documentclass[aip, amsmath, amssymb, reprint]{revtex4-1}

\usepackage{graphicx}
\usepackage{dcolumn}
\usepackage{bm}
\usepackage[utf8]{inputenc}
\usepackage[T1]{fontenc}
\usepackage{mathptmx}
\usepackage{etoolbox}
\usepackage{natbib,hyperref}
\usepackage[percent]{overpic}
\usepackage[dvipsnames]{xcolor}
\usepackage{soul}
\graphicspath{{./fig_low_res/}{./fig/}{./}}
\newcommand{\highlight}[1]{#1} 

\newcommand{\EQ}{\begin{equation}}
\newcommand{\EN}{\end{equation}}
\newcommand{\EQA}{\begin{eqnarray}}
\newcommand{\ENA}{\end{eqnarray}}

\newcommand{\Fig}[1]{Figure~\ref{#1}}

{}
{}
{}

{}
{}
{}
{}
{}
{}
{}
{}
{}
{}
{}
{}
{}
{}
{}
{}
{}
{}
{}
{}
{}

{}
{}
{}

{}

{}
{}

\newcommand{\rr}{\bm{r}}
\newcommand{\kk}{\bm{k}}

\newcommand{\xx}{\bm{x}}

\newcommand{\BB}{\bm{B}}

\newcommand{\uu}{\bm{u}}

\newcommand{\JJ}{\bm{J}}

\newcommand{\AAA}{\bm{A}}

\newcommand{\FF}{\bm{F}}

\newcommand{\nab}{\mbox{\boldmath $\nabla$} {}}

\newcommand{\SSSS}{\mbox{\boldmath ${\sf S}$} {}}

\newcommand{\DD}{{\rm D} {}}

\newcommand{\dd}{{\rm d} {}}

\def\cs{c_{\rm s}}

\newcommand{\includefigure}[3]{
\sethlcolor{white}
\begin{overpic}[width=#2\columnwidth]{#1}
 \put (5,5) {\large \hl{#3}}
\end{overpic}
}

\makeatletter
\def\@email#1#2{%
 \endgroup
 \patchcmd{\titleblock@produce}
  {\frontmatter@RRAPformat}
  {\frontmatter@RRAPformat{\produce@RRAP{*#1\href{mailto:#2}{#2}}}\frontmatter@RRAPformat}
  {}{}
}%
\makeatother
\begin{document}

\preprint{AIP/123-QED}

\title[Twisted Magnetic Knots and Links]{Twisted Magnetic Knots and Links}

\author{S. Candelaresi}
\email{Simon.Candelaresi@gmail.com.}
\affiliation{
High-Performance Computing Center, Nobelstrasse 19, Stuttgart 70569, Germany
}
\affiliation{
School of Mathematics and Statistics, University of Glasgow, Glasgow G12 8QQ, United Kingdom
}
\author{C. Beck}
\affiliation{ 
School of Mathematics and Statistics, University of Glasgow, Glasgow G12 8QQ, United Kingdom
}

\date{\today}

\begin{abstract}
For magnetic knots and links in plasmas we introduce an internal twist
and study their dynamical behavior in numerical simulations.
We use a set of helical and non-helical configurations
and add an internal twist that cancels the helicity of the helical
configurations or makes a non-helical system helical.
These fields are then left to relax in a magnetohydrodynamic
environment.
In line with previous works we confirm the importance of magnetic helicity
in field relaxation.
However, an internal twist, as could be observed in coronal magnetic loops,
does not just add or subtract helicity, but also introduces
an alignment of the magnetic field with the electric current, which is
the source term for helicity.
This source term is strong enough to lead to a significant change
of magnetic helicity, which for some cases leads to a loss of the stabilizing
properties expressed in the realizability condition.
Even a relatively weak internal twist in these magnetic fields leads to
a strong enough source term for magnetic helicity that
for various cases even
in a low diffusion environment
we observe an inversion in sign of magnetic helicity
within time scales much shorter than the diffusion time.
We conclude that in solar and stellar fields an internal twist does not
automatically result in a structurally stable configuration
and that the alignment of the magnetic field with the electric current
must be taken into account.
\end{abstract}

\maketitle

\section{Introduction}

The majority (over $99$\%) of the observable matter in the universe
is known to be in a state of plasma.
Being a plasma, the electric conductivity is generally high
and we observe electric currents and the magnetic
fields these currents generate.
These magnetic fields have been observed in planets, our Sun, stars,
magnetars and galaxies.

Through the Lorentz force the magnetic field acts on the medium.
At the same time, the motions of the medium leads to a change
in electric current and the magnetic field.
This non-linear interaction gives rise to various physically
relevant effects.
For instance, it can be used to explain the exponential amplification
of a weak magnetic field (dynamo effect) in stars and galaxies
(see e.g.\ \onlinecite{Krause-Radler-1971-SolMagFields, BrandenbSubramanianReview2005}).
Other phenomena for which this interaction is crucial are solar
coronal mass ejections
(e.g.\ \onlinecite{Barnes-2007-670-L53-ApJL})
and stellar flares
(e.g.\ \onlinecite{Schrijver2009AdSpR43}).

For laboratory and astrophysical plasmas we know that
magnetic fields and electric currents have a significant effect on the
dynamics and evolution of the system
(e.g.\ \onlinecite{MoffattBook1978,
Biskamp-2003-Magnetohydrodynamic_Turbulence, Priest-2014-MHDSun}).
This is particularly true for magnetically dominated regions like
the solar corona \citep{Kasper-Klein-2021-127-8-PRL} where
the magnetic pressure is larger than the hydrostatic pressure
and magnetic forces dominate.
In laboratory devices, the magnetic field is particularly
important in confining hot plasmas in fusion devices like
tokamaks or the Large Helical Device in Toki \citep{Motojima2006LHD}.

During solar maxima we can observe large-scale magnetic flux tubes
on the solar surface (e.g.\ \onlinecite{Ballegooijen-2004-612-519-ApJ}).
Those are a result of the dynamo action within the Sun
which is responsible for the 11 year solar cycle.
While the plasma motions give rise to particular magnetic field
structure, the magnetic field in turn affects the plasma motions.
Here, the geometry of the field lines plays an important role in
the dynamics.
For instance, a strong magnetic flux tube will rise due to
magnetic buoyancy and a high curvature in the field lines gives rise
to additional Lorentz forces.
\highlight{
Finally, reconnection events near magnetic null points
\citep{Filippov-1999-185-2-SolPhys}
(where the magnetic field vanishes)
give rise to strong particle acceleration
\cite{Pontin-Hornig-2005-99-77-GApFD}.
}

While it should be evident that the field's geometry affects the dynamics,
it is less evident that its topology, i.e.\ field line connectivity, has
a strong effect.
The most used quantifier for the magnetic field topology is the
magnetic helicity, which measures the linkage, twisting, braiding
and knottedness of the magnetic field \citep{MoffattBook1978}.
For astrophysical parameter regimes we know that this quantity
changes on much longer time scales than the dynamics of the system.
For instance, we have 11 years for the solar magnetic cycle, which compares
to millions of years for the magnetic helicity decay time
\highlight{
(e.g. \onlinecite{BrandenbSubramanianReview2005}).
}

This conservation has dramatic consequences for the plasma dynamics.
For a freely relaxing field \onlinecite{Woltjer-1958-489-91-PNAS} derived a
minimum energy state with the restriction of magnetic helicity conservation.
This state has the form of a linear force-free field with vanishing Lorentz force.
Calculations by \onlinecite{Taylor-1974-PrlE} and \onlinecite{Taylor1986}
in the fusion plasma context showed that a minimum energy state
is a non-linear force-free state.
\onlinecite{ArnoldHopf1986} then demonstrated that the magnetic energy
is bound from below by the presence of magnetic helicity,
the so called realizability condition.
This puts severe restrictions on the final equilibrium state,
and the intermediate states the system is allowed to go through
during its evolution.

\highlight{
Magnetic helicity is also being exploited in turbulent dynamo theory.
In a turbulent plasma it has been shown (e.g.\
\onlinecite{Frisch-Pouquet-Leorat-1975-JFluidMech, KleeorinRuzmaikin82, alpha2_periodic13})
that helicity conservation leads to an inverse cascade
where small-scale magnetic energy is transformed into large-scale energy.
As a result, an initially weak small-scale magnetic field is being
transformed into a strong large-scale field.
This effect has been used to explain the existence of strong large-scale
magnetic fields in planets, stars and galaxies.
}

Following these early analytical results and with the rise of computing power,
various numerical calculations on magnetic field relaxation have been performed.
Ideal magnetohydrodynamics (MHD) simulations that preserve the field's topology
were proposed by \onlinecite{Craig-Sneyd-1986-311-451-ApJ} and used
by \onlinecite{Craig-Sneyd-1990-357-653-ApJ} to study the stability of coronal
flux tubes in high magnetic Reynolds number regimes.
Using resistive MHD simulations of relaxing magnetic braids
\onlinecite{Yeates_Topology_2010} demonstrated that the relaxed state proposed by
\onlinecite{Taylor-1974-PrlE} is not necessarily reached, particularly in presence of
further topological invariants.
Further works on resistive magnetic field relaxation by \onlinecite{fluxRings10}
and \onlinecite{knotsDecay11} showed the restricting effect of the realizability
condition, but also highlighted that magnetic fields with no helicity
that are topologically non-trivial exhibit a behavior reminiscent of
a restricted relaxation.

\highlight{
Magnetic helicity is a second order invariant of MHD.
Higher order invariants can be defined, like those by
\onlinecite{Ruzmaikin-Akhmetiev-1994-331-1-PhysPlasm}.
They introduced third and fourth order invariants.
However, in presence of even a small amount of magnetic resistivity,
magnetic field line reconnection destroys these invariants which
makes them less suitable for studying real plasmas.
Other fourth order invariants are the quadratic helicities
(e.g.\
\onlinecite{Akhmetev-2012-278-10-ProcSteklov,
helicity2,
Akhmetev-Candelaresi-2018-84-6-JPlasmPh}
).
Those have been shown to pose certain restrictions to the
plasma dynamics, similar to the helicity.
}

Since authors have almost exclusively focused on the magnetic helicity as
topological quantifier, there is little work on topologically non-trivial
non-helical fields.
Furthermore, the premise that magnetic helicity is conserved within
dynamical time scales has often been taken for granted in high Reynolds number
regimes.
However, the rate of change for the helicity also depends on the alignment
of the electric current density with the magnetic field which can
be significant in some environments.
Here we will present a series of topologically non-trivial magnetic field
configuration that undergo a resistive relaxation using the MHD framework.
Those consist of knots and rings and can be helical or non-helical.
We also add an internal twist to the flux tubes that can add to the helicity
content or cancel it.

\section{Methods}

\subsection{Knot Construction}

We construct a set of magnetic knots of which some have an internal twist.
Starting at the central spine of the knot we then construct the full tube
with finite width,
\highlight{
and leave the space outside these tubes with no magnetic field.
}
The equation describing their center line is given as
\EQ\label{eq: knot}
\xx(s) =
\left(
\begin{array}{c}
(C + \sin(s n_{\rm f}))\sin[s(n_{\rm f} -1)] \\
(C + \sin(s n_{\rm f}))\cos[s(n_{\rm f} -1)] \\
D\cos(s n_{\rm f})
\end{array}
\right),
\EN
where $C$ specifies the knot size in the $xy$-plane,
$D$ the extension in the $z$-direction,
$s \in [0, 2\pi]$ is the knot parameter and
$n_{\rm f}$ the number of foils of the knot
($n_{\rm f} = 3$ for a trefoil knot).
Here we choose $n_{\rm f} \in \{3, 4, 5\}$,
$C = 2$ and $D = 2.5$.
Only for the trivial ring we choose $D = 0$ and $n_{\rm f} = 0$.
As width for the flux tubes we choose $0.5$.

The magnetic field $\BB$ is tangent to the central spine.
So, we can write
\EQ
\hat{\BB} = \xx'(s)/|\xx'(s)|.
\EN
In order to construct a space filling field $\BB$ for every position vector
$\rr$ in our domain we find the closest point on the spine to $\rr$.
This will give us the minimal curve parameter $s$ and with $\xx'(s)$
it gives us $\hat{\BB}$ at the point $\rr$.
To find the appropriate magnetic field strength at this position
we need to take into account distance to the curve, which can be easily computed
as $|\rr - \xx(s)|$.
Using a field strength profile we obtain the preliminary strength.
In order to satisfy $\nab\cdot\BB = 0$ we need to take into account the curvature, which
we can easily find as $\xx''(s)$.

\begin{figure}[t!]\begin{center}
\includegraphics[width=\columnwidth]{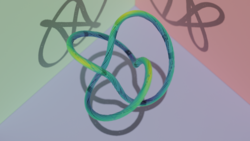}
\end{center}
\caption[]{
Magnetic streamlines for the twisted trefoil knot with helicity at time $0$.
}\label{fig: streamlines_t0_n3_r256_tw1.219755}
\end{figure}

So far we have constructed a magnetic knot with finite width, but no internal twist.
By adding a component that is not tangent to $\xx(s)$ we can
add such a twist to the field.
An example magnetic trefoil knot is shown in \Fig{fig: streamlines_t0_n3_r256_tw1.219755},
where we chose $n_{\rm f} = 3$ and a twist parameter of $1.22$.

Apart from the class of knots described in equation \eqref{eq: knot} we also
study one that requires a different description, the $8_{18}$ knot.
Its central spine is given as
\EQ\label{eq: iucaa}
\xx(s) =
\left(
\begin{array}{c}
(C + \sin(4s))\sin(3s) \\
(C + \sin(4s))\cos(3s) \\
D\cos(8s)
\end{array}
\right),
\EN
with the same parameters as in equation \eqref{eq: knot}.
A rendering of the $8_{18}$ knot can be seen in \Fig{fig: streamlines_IUCAA}
\highlight{
(panel a)
}.

Apart from knots we also investigate the relaxation of links
that are topologically non-trivial.
One of them is the Borromean rings configuration.
Unlike in its usual depiction, we do not project it onto a plane and
choose a more symmetric representation consisting of
three magnetic flux ovals.
Those are arranged such that by removing one, the remaining two
\highlight{
constitute two unlinked and topologically trivial flux rings.
}
A rendering of the Borromean rings can be seen in \Fig{fig: streamlines_borromean}.

The last configuration consists of one flux ring that is linked
with two others on opposing sides, like a short chain where
the flux rings are the links.
We can think of this configuration as a sort of double Hopf link
where each of the two links contribute to the magnetic helicity content.
By changing the sign of the flux in one of the outer rings
we can construct topologically different cases, one with helicity
and one without.
A rendering of the triple ring configurations rings can be seen in
\Fig{fig: streamlines_linked_tori_h0}.

\highlight{
Since our computational code uses the magnetic vector potential $\AAA$ instead
of the magnetic field $\BB = \nab\times\AAA$ for the computation, we need
as last step of the construction to compute the inverse curl.
We can do this by first taking the curl
\EQ
\JJ = \nab\times\BB,
\EN
which gives us the current density $\JJ$.
We then express the current density using the magnetic vector potential as
\EQA
\JJ & = & \nab\times\nab\times\AAA \nonumber \\
 & = & -\Delta \AAA + \nab (\nab\cdot\AAA).
\ENA
Using the Coulomb gauge $\nab\cdot\AAA = 0$ we can then perform a Fourier
transform on the problem
\EQ
{\cal F}\{\JJ\}(\kk) = -k^2 {\cal F}\{ \AAA \}(\kk),
\EN
invert it in $k$-space
\EQ
{\cal F}\{ \AAA \}(\kk) = -{\cal F}\{ \JJ \}(\kk) /k^2,
\EN
solve this algebraic equation and transform this solution back into real space.
Since all of our initial conditions are $0$ at the domain boundaries
we have periodic initial conditions and the Fourier method leads to
valid initial vector potentials $\AAA$, such that $\nab\times\AAA$
matches our initial construction of $\BB$, up to small numerical errors.
}

\subsection{Simulation Setup}

The magnetic knot or link is the initial condition of our numerical experiments.
We place the configuration into a Cartesian box
with size $[8, 8, 8]$.
Being periodic, there are no fluxes outside the domain.

In our model we make use of the magnetohydrodynamics equations
of a viscous, diffusive, compressible gas
\EQA
\frac{\partial \AAA}{\partial t} & = & \uu\times\BB + \eta\nab^2\AAA,
\label{eq: induction} \\
\frac{\DD \uu}{\DD t} & = & -\cs^{2}\nab \ln{\rho} + \frac{\JJ\times\BB}{\rho} + \FF_{\rm visc},
\label{eq: momentum} \\
\frac{\DD \ln{\rho}}{\DD t} & = & -\nab \cdot \uu,
\label{eq: continuity}
\ENA
with the fluid velocity $\uu$,
constant magnetic resistivity (diffusivity) $\eta$,
advective derivative $\DD / \DD t = \partial/\partial t + \uu\cdot\nab$,
sound speed $\cs = \gamma p/\rho$,
density $\rho$,
electric current density $\JJ = \nab\times\BB$ and
viscous force $\FF_{\rm visc}$.
The viscous force captures the friction between molecules and
is given as $\FF_{\rm visc} = \rho^{-1}\nab\cdot 2\nu\rho\SSSS$,
with the traceless rate of strain tensor $S_{ij} = \frac{1}{2}(u_{i,j} + u_{j,i}) - \frac{1}{3}\delta_{ij}\nab\cdot\uu$.
We make implicitly use of an equation of state of an ideal monatomic gas, as we eliminated pressure $p$
for density $\rho$.
To solve the equations we make use of the {\sc PencilCode}
\cite{Brandenburg-2020-Zenodo, PencilCodeCollaboration-2021-58-2807-JOpSoSoft}
(\url{https://github.com/pencil-code}),
which is a sixth order in space and fourth order in time
finite difference code.

To keep kinetic and magnetic dissipation low we choose for the viscosity
and magnetic resistivity $\nu = \eta = 10^{-3}$
for most of our simulations.
That way, the magnetic helicity
\highlight{
won't be significantly affected by magnetic diffusion
}
during dynamically relevant times.
In a comparison we will also show a few simulations using
$\nu = \eta = 5\times 10^{-4}$.

\highlight{
For the boundary conditions we choose periodic boundaries for all of our setups.
Not only is this a computationally unproblematic condition, but it helps us
conserving energy (magnetic and kinetic) and magnetic helicity together
with current helicity.
}

\subsection{Magnetic Helicity Dissipation}

In an ideal fluid with no viscosity and no magnetic diffusion ($\nu = \eta = 0$)
the magnetic helicity
\EQ
H_{\rm m} = \int_V \AAA\cdot\BB \ \dd V
\EN
is exactly conserved.
$\nu$ and $\eta$ are finite, albeit very small, for physically interesting objects,
like stars, galaxies, planetary interiors and fusion devices.
From the MHD equations \eqref{eq: induction}--\eqref{eq: continuity} we
can easily derive the rate of change of the magnetic helicity as
\EQ
\label{eq: helicity_dissipation}
\frac{\partial H_{\rm m}}{\partial t} = -2\eta\int_{V}\JJ\cdot\BB\ \dd V.
\EN

Here we see that for $\eta = 0$ magnetic helicity is conserved.
For a finite $\eta$ the rate of dissipation depends on the magnetic field strength,
the current density and their alignment.
In turbulent media the current can reach relatively high values
as the magnetic field undergoes repeated reconnection events.
So, even with a small $\eta$ we cannot exclude a significant change
in the magnetic helicity content.
Due to its importance in helicity conservation, we will monitor this
value for our simulations.

\section{Results}

\subsection{Geometry Evolution}

Before analyzing the evolution of the magnetic field and helicity in time
we study the geometric evolution of the field lines.
For the visualization of the magnetic streamlines we make use
of the open-source tool BlenDaViz \footnote{https://github.com/SimonCan/BlenDaViz}
and trace a number of magnetic streamlines.
The streamlines have their seeds where the field strength is relatively strong,
so we avoid noise generated in weak field domains.
Here we compare the topological structure of a selection of magnetic fields
at their initial configuration and final state when we stop the calculations.

\highlight{
For all of our calculations we observe a fast dynamical regime until a
simulation time of less than $100$.
This regime is dominated by the Lorentz force that causes some of the
configurations to contract.
As the flux tubes approach, large currents are being generated which then
leads to energy loss through Ohmic heating.
After this regime, the resulting magnetic field is subject to a slow
diffusion of time scale of $2000$ code time units.
}

\subsubsection{Knots}

\begin{figure*}[t!]
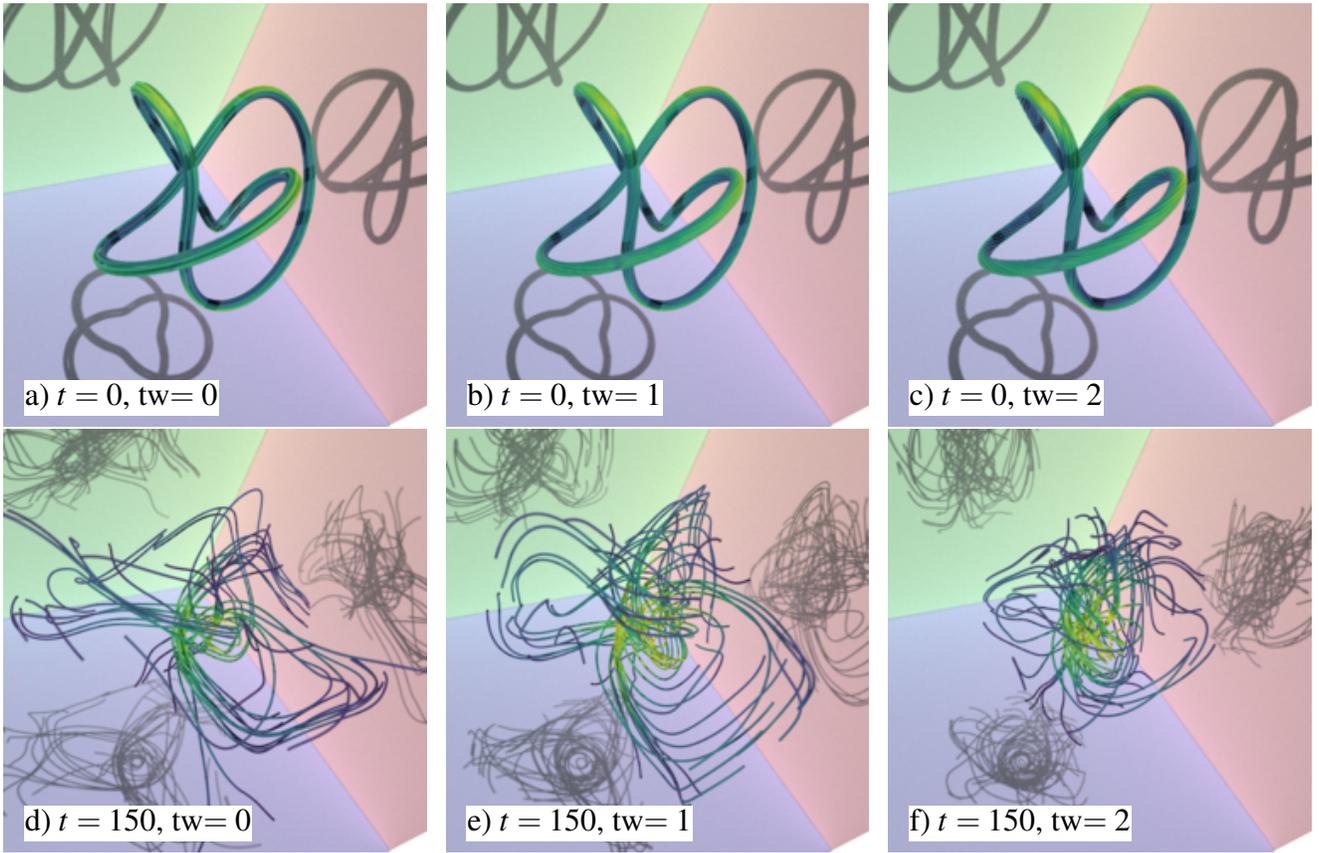
\begin{center}
\includefigure{n3_r256_tw0_t0}{0.65}{a) $t=0$, tw$=0$}
\includefigure{n3_r256_tw061009_t0}{0.65}{b) $t=0$, tw$=1$}
\includefigure{n3_r256_tw1_219755_t0}{0.65}{c) $t=0$, tw$=2$} \\
\includefigure{n3_r256_tw0_tf}{0.65}{d) $t=150$, tw$=0$}
\includefigure{n3_r256_tw061009_tf}{0.65}{e) $t=150$, tw$=1$}
\includefigure{n3_r256_tw1_219755_tf}{0.65}{f) $t=150$, tw$=2$}
\end{center}
\caption[]{
Initial magnetic streamlines for the trefoil knot configuration (top) using
different twist parameter ${\rm tw}$ such that the left has no twist,
the middle has a twist that reduces the helicity to zero and the
right a strong twist with opposite helicity to the left one.
The lower figures show the streamlines at time $150$.
}\label{fig: streamlines_trefoil}
\end{figure*}

In \Fig{fig: streamlines_trefoil} we show the initial field configuration
for the trefoil knots together with the final field configurations
after magnetic relaxation.
For a fair comparison the final time is $t = 150$ in code units for all
configurations.
The internal twist is ascending from left to right and is chosen such
that the second knot
\highlight{
(panels b and e)
}
has a twist that generates a helicity exactly
opposite of the helicity from the knotedness.
The highly twisted knot has a helicity of equal strength,
but opposite sign of the untwisted knot
\highlight{
(panels c and f)
}
.

From the magnetic helicity conservation and the realizability
condition we would expect that after relaxation, and the accompanying
field line reconnection, the two helical cases
\highlight{
(a and c)
}
should approximately
stay helical while the non-helical case should lose all its initial structure.
However, all three configurations clearly show a twist and with that a helical field
at final times.
This is on contrast to previous findings where non-helical fields
would evolve into a state with no discernible structure \cite{fluxRings10}.
This is clearly due to a generation of magnetic helicity within dynamical
time scales, as we will
discuss in more detail further below.
This helicity generation is clearly not strong enough for the untwisted
knot to be transformed into a non-helical field, which would then
lose its initial structure.
But it is fast enough for the twisted knot to generate a large enough
amount of helicity, in a time much shorter than the diffusive time,
so that the final structure is similarly twisted as the plain knot.

For the $4$-foil and $5$-foil knot configurations we observe
qualitatively the same behaviour.
Since the initial helicity is larger than for the trefoil knot
the final state is more twisted.
For the non-helical case with the opposing internal twist,
we observe a significant generation of twist and therefore helicity.
To conserve space we moved the results for the $4$-foil
and $5$-foil knots to appendix \ref{app: field_evolution}.

\subsubsection{Triple Rings}

For the triple rings configurations we have five realizations
divided into two sets.
One set derives from the non-helical configuration and
the other form the helical.
By adding an internal twist to the non-helical we can induce
a magnetic helicity content of the same size as the untwisted
helical triple ring.
For the helical triple rings we can add a twist so that the helicity
exactly vanishes and an opposite twist to double the magnetic helicity
compared to the untwisted case.

\begin{figure*}[t!]
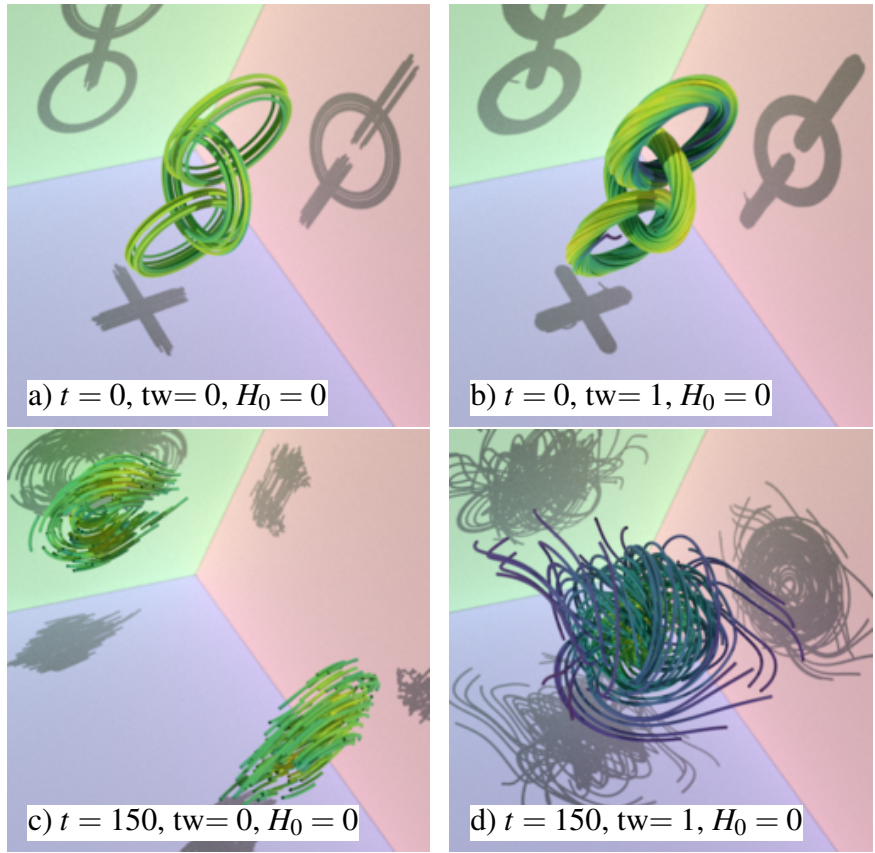
\begin{center}
\includefigure{linked_tori_h0_r256_tw0_t0}{0.65}{a) $t=0$, tw$=0$, $H_0=0$}
\includefigure{linked_tori_h0_r256_tw0_271_t0}{0.65}{b) $t=0$, tw$=1$, $H_0=0$} \\
\includefigure{linked_tori_h0_r256_tw0_tf}{0.65}{c) $t=150$, tw$=0$, $H_0=0$}
\includefigure{linked_tori_h0_r256_tw0_271_tf}{0.65}{d) $t=150$, tw$=1$, $H_0=0$}
\end{center}
\caption[]{
Initial magnetic streamlines for the triple ring configurations (top).
With twist parameter ${\rm tw} = 0$ this would be a non-helical field
\highlight{
(hence $H_0 = 0$)
}
.
The lower figures show the streamlines at time $150$.
}\label{fig: streamlines_linked_tori_h0}
\end{figure*}

\begin{figure*}[t!]
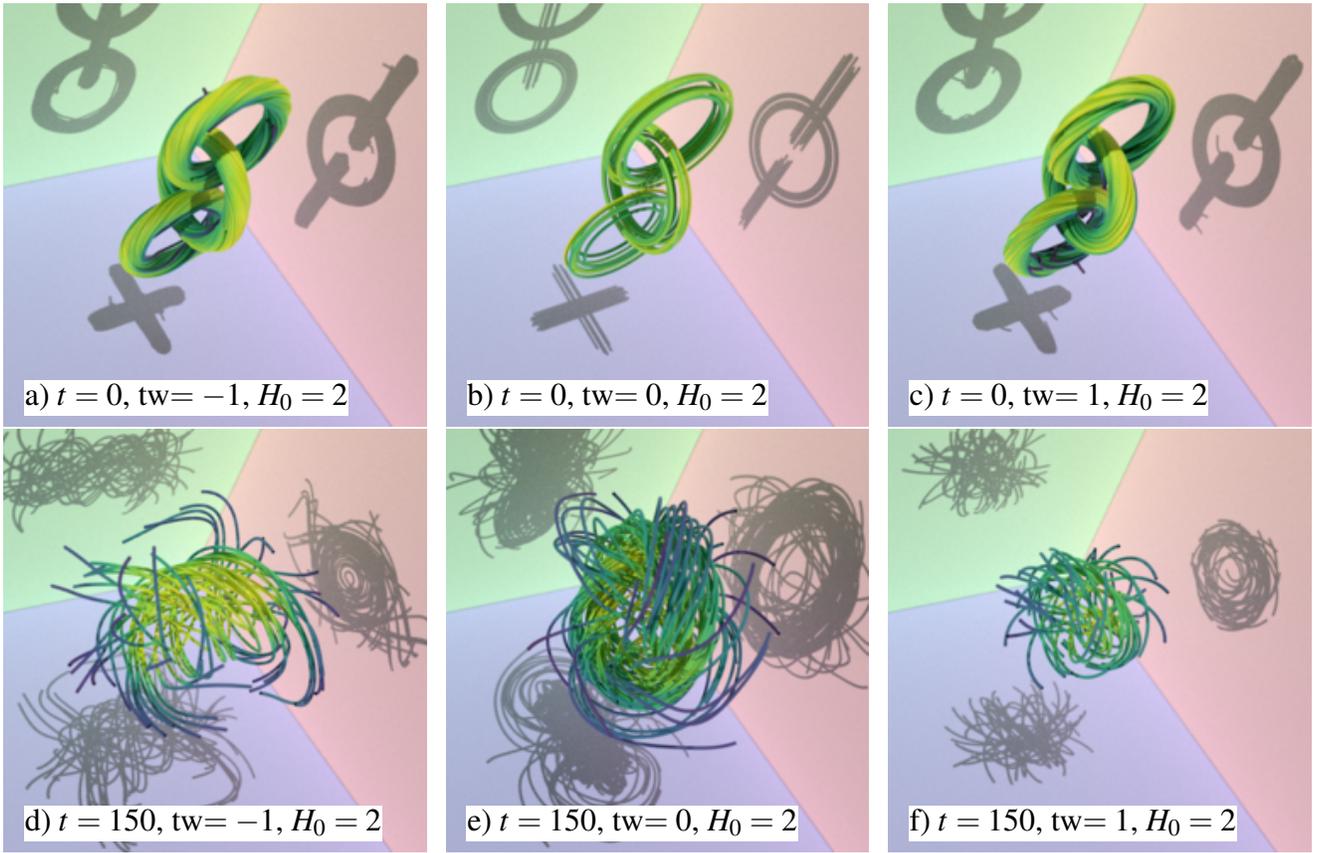
\begin{center}
\includefigure{linked_tori_h2_r256_tw-0_271_t0}{0.65}{a) $t=0$, tw$=-1$, $H_0=2$}
\includefigure{linked_tori_h2_r256_tw0_t0}{0.65}{b) $t=0$, tw$=0$, $H_0=2$}
\includefigure{linked_tori_h2_r256_tw0_271_t0}{0.65}{c) $t=0$, tw$=1$, $H_0=2$} \\
\includefigure{linked_tori_h2_r256_tw-0_271_tf}{0.65}{d) $t=150$, tw$=-1$, $H_0=2$}
\includefigure{linked_tori_h2_r256_tw0_tf}{0.65}{e) $t=150$, tw$=0$, $H_0=2$}
\includefigure{linked_tori_h2_r256_tw0_271_tf}{0.65}{f) $t=150$, tw$=1$, $H_0=2$}
\end{center}
\caption[]{
Initial magnetic streamlines for the triple ring configurations (top).
With twist parameter ${\rm tw} = 0$ this would be a helical field
\highlight{
(hence $H_0 = 2$)
}
.
The left field has negative twist parameter and the right one positive.
With the amount of negative twist the magnetic helicity of the left
configuration is zero and the helicity of the right configuration
is double to the untwisted case.
The lower figures show the streamlines at time $150$.
}\label{fig: streamlines_linked_tori_h2}
\end{figure*}

The three linked rings, by contrast to the knots, do not show much unexpected behaviour.
Here, the helical configurations retain their topological structure
during the relaxation, while the non-helical ones decay into
a field that is more trivial (see \Fig{fig: streamlines_linked_tori_h0}
and \Fig{fig: streamlines_linked_tori_h2}).
Unlike the twisted non-helical knots, the twisted non-helical
triple ring field (\Fig{fig: streamlines_linked_tori_h2}, \highlight{panels a and c})
does not show
any significant generation of helicity.
With that the relaxed field is roughly trivial.

We attribute this different behavior to the absence of any significant
magnetic helicity generation from the internal twist.
\highlight{
Currently we cannot explain why this is the case for the other test configurations,
but not for the triple rings configuration.
}

\subsubsection{Borromean Rings}

\begin{figure*}[t!]
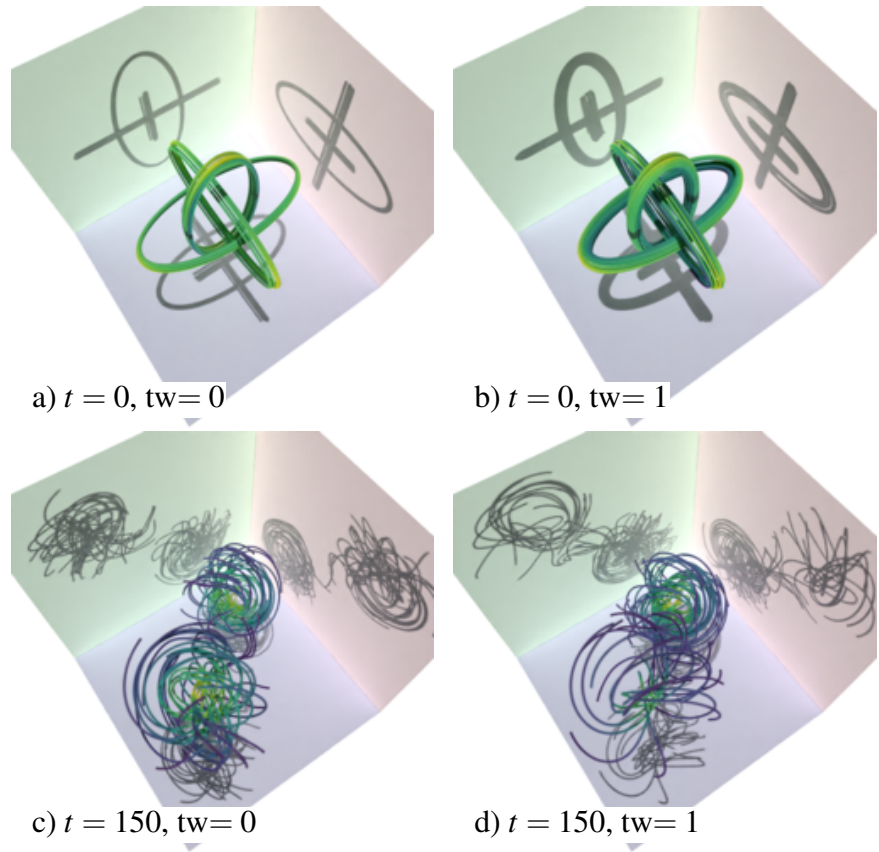
\begin{center}
\includefigure{borromean_r256_tw0_t0}{0.65}{a) $t=0$, tw$=0$}
\includefigure{borromean_r256_tw0_0351_t0}{0.65}{b) $t=0$, tw$=1$} \\
\includefigure{borromean_r256_tw0_tf}{0.65}{c) $t=150$, tw$=0$}
\includefigure{borromean_r256_tw0_0351_tf}{0.65}{d) $t=150$, tw$=1$}
\end{center}
\caption[]{
Initial magnetic streamlines for the Borromean rings configurations (top)
with (left) and without (right) internal twist.
The lower figures show the streamlines at time $150$.
\highlight{
Note the slightly changed perspective compared to the other streamline plots.
}
}\label{fig: streamlines_borromean}
\end{figure*}

For the untwisted Borromean rings we observe the same behaviour
as in the work by \onlinecite{knotsDecay11}.
This is not surprising, as the simulation setup is very close to theirs
and we make use of the same numerical code.
Here we see that the non-helical Borromean rings undergo a series
of reconnection events and relax into a configuration consisting of
two separate and oppositely twisted fields (\Fig{fig: streamlines_borromean},
\highlight{panel c}).

The Borromean rings configuration with the internal twist, however,
relaxes into a relatively compact field
\highlight{
(panel d)
}
.
This field has a clear twist, which is a consequence of the magnetic
helicity conservation in this low resistivity system.
Like the triple-ring, and in contrast to the knots, we observe
a behavior predicted from the magnetic helicity conservation.

\subsubsection{$8_{18}$ Knot}

\begin{figure*}[t!]
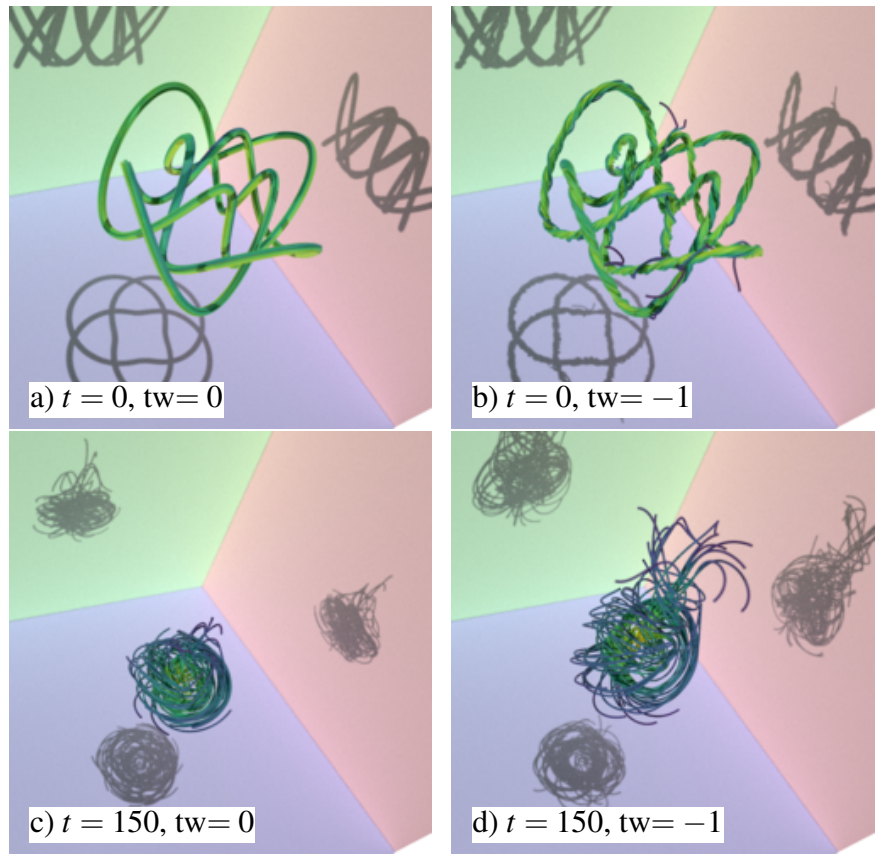
\begin{center}
\includefigure{iucaa_r256_tw0_t0}{0.65}{a) $t=0$, tw$=0$}
\includefigure{iucaa_r256_tw-4_t0}{0.65}{b) $t=0$, tw$=-1$} \\
\includefigure{iucaa_r256_tw0_tf}{0.65}{c) $t=150$, tw$=0$}
\includefigure{iucaa_r256_tw-4_tf}{0.65}{d) $t=150$, tw$=-1$}
\end{center}
\caption[]{
Initial magnetic streamlines for the $8_{18}$ knot configurations (top)
with and without internal twist.
The lower figures show the streamlines at time $150$.
}\label{fig: streamlines_IUCAA}
\end{figure*}

The plain $8_{18}$ knot is a non-helical configuration.
With helicity conservation we expect a topologically trivial relaxed field
at final simulation time.
Although we indeed do not observe any significant change in magnetic helicity
the final state appears to be a topologically non-trivial
bundle of twisted field lines (\Fig{fig: streamlines_IUCAA}
\highlight{
panel c}).
\highlight{
However, upon closer inspection, we see that this bundle has only
a small net twist, which makes it topologically trivial.
}

\highlight{
The twisted $8_{18}$ knot is a helical field with its helicity coming entirely
from the twist.
If its magnetic helicity is approximately conserved during its relaxation,
we should see a topologically non-trivial field at the end of the simulation.
This is indeed what we observe in panel d, where there is a clear twist
at the final time.
}

\subsection{Magnetic Helicity}

In the limit of vanishing magnetic resistivity, we know that
the magnetic helicity is conserved (see equation \eqref{eq: helicity_dissipation}).
However, this is only true if the $\JJ\cdot\BB$ term is
not so large as to compensate for a small $\eta$.
If there is a net alignment between the magnetic field $\BB$ and
electric current density $\JJ$ then we should see dissipation.
For a net anti-alignment we would observe an increase in helicity.
Potential loci for this to happen are reconnection regions,
since there $\JJ$ and $\BB$ are relatively aligned.
Furthermore, the current in these regions typically increases significantly
which leads to Ohmic heating.

\highlight{
Other sources of $\JJ$--$\BB$ alignment are magnetic twist.
For an untwisted magnetic flux tube the generating currents
are purely toroidal.
Adding a twist has two effects.
First, the toroidal magnetic field component is now aligned with
the toroidal current that generates the untwisted part of the
magnetic field.
Second, the twist component is generated by an axial magnetic field
and is aligned with the axial magnetic field of the untwisted
magnetic field component.
}

\highlight{
To better understand the helicity evolution for our simulations
we monitor the volume averaged magnetic helicity in time.
The definition of this average is
\EQ
\langle \AAA\cdot\BB \rangle = \frac{1}{V} \int_{V} \AAA\cdot\BB \ \dd V.
\EN
Since we are also interested in the generation of helicity we also monitor
the spatially averaged current helicity
\EQ
\langle \JJ\cdot\BB \rangle = \frac{1}{V} \int_{V} \JJ\cdot\BB \ \dd V.
\EN
}

\subsubsection{Knots}

\begin{figure}[t!]\begin{center}
\includegraphics[width=\columnwidth]{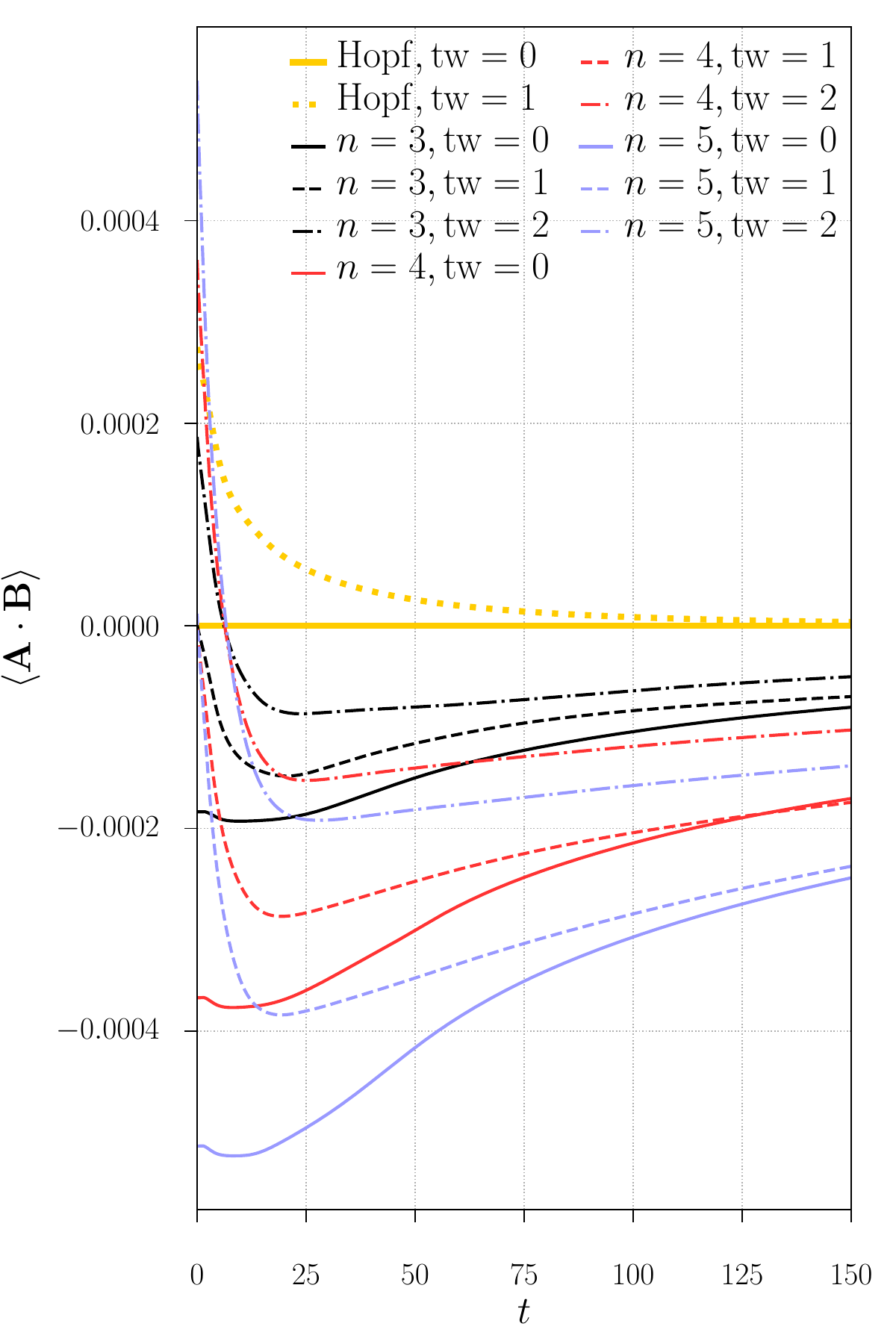}
\end{center}
\caption[]{
Mean magnetic helicity in dependence of time for the knot configurations
\highlight{and the Hop link} using
different twist parameters ${\rm tw}$.
}\label{fig: abm_knots}
\end{figure}

For the untwisted knots we observe
a relatively slow loss of helicity (\Fig{fig: abm_knots}, solid lines).
While the initial loss is fast, after some short time it slows down.
Nevertheless, the loss is faster than the expected resistive decay
time of $\tau_{\rm res} = L^2/\eta \approx 1000$, where $L \approx 1$ is a
typical length scale of the system.
So we already see that even without internal twist, the contribution
of the $\JJ$--$\BB$-alignment is significant.

For the twisted knots with zero helicity (\Fig{fig: abm_knots}, dashed lines)
we see a significant generation of negative helicity at initial times.
The time scales are much shorter than the diffusion time.
This change is again due to the net alignment of $\JJ$ and $\BB$, which
is even more significant than for the untwisted knots.

For the highly twisted knots with opposite magnetic helicity we see a strong
initial decrease of helicity.
Similar to the previous cases, this is due to an even stronger alignment of $\JJ$
and $\BB$.
Within this very short time period we even observe a sign change and
an approach to the magnetic helicity of the untwisted case.

\begin{figure}[t!]\begin{center}
\includegraphics[width=\columnwidth]{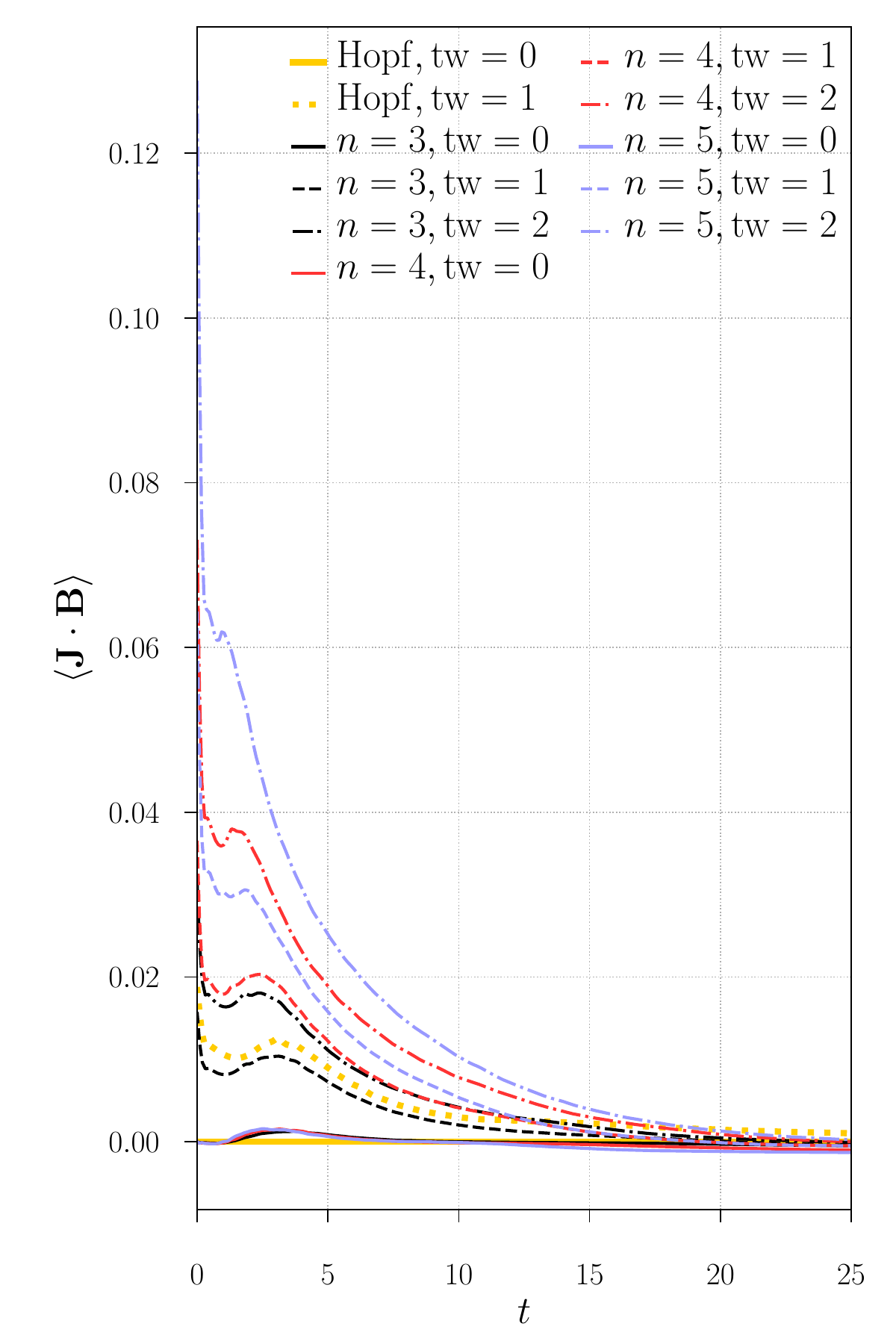}
\end{center}
\caption[]{
Mean current helicity in dependence of time for the knot configurations
\highlight{and the Hop link} using
different twist parameters ${\rm tw}$.
}\label{fig: jbm_knots}
\end{figure}

The $\JJ\cdot\BB$ term is so dominant, that within much less than the diffusion time
we see a significant decrease in magnetic helicity for the twisted cases.
This effect is increases with twist parameter (see \Fig{fig: jbm_knots}).

For comparison we also plot the Hopf link in the same graph.
This configuration is helical due to the mutual linking of two magnetic
flux rings.
Similarly to the knots we can add an internal twist to the magnetic flux tubes
to reduce the helicity content to $0$.
Unlike the twisted knots with zero initial helicity, the twisted
Hopf link with no initial helicity does not show any significant change within
dynamical times.
This shows that a twist does not necessarily lead to a $\JJ$--$\BB$ alignment.

\subsubsection{Triple Rings}

\begin{figure}[t!]\begin{center}
\includegraphics[width=\columnwidth]{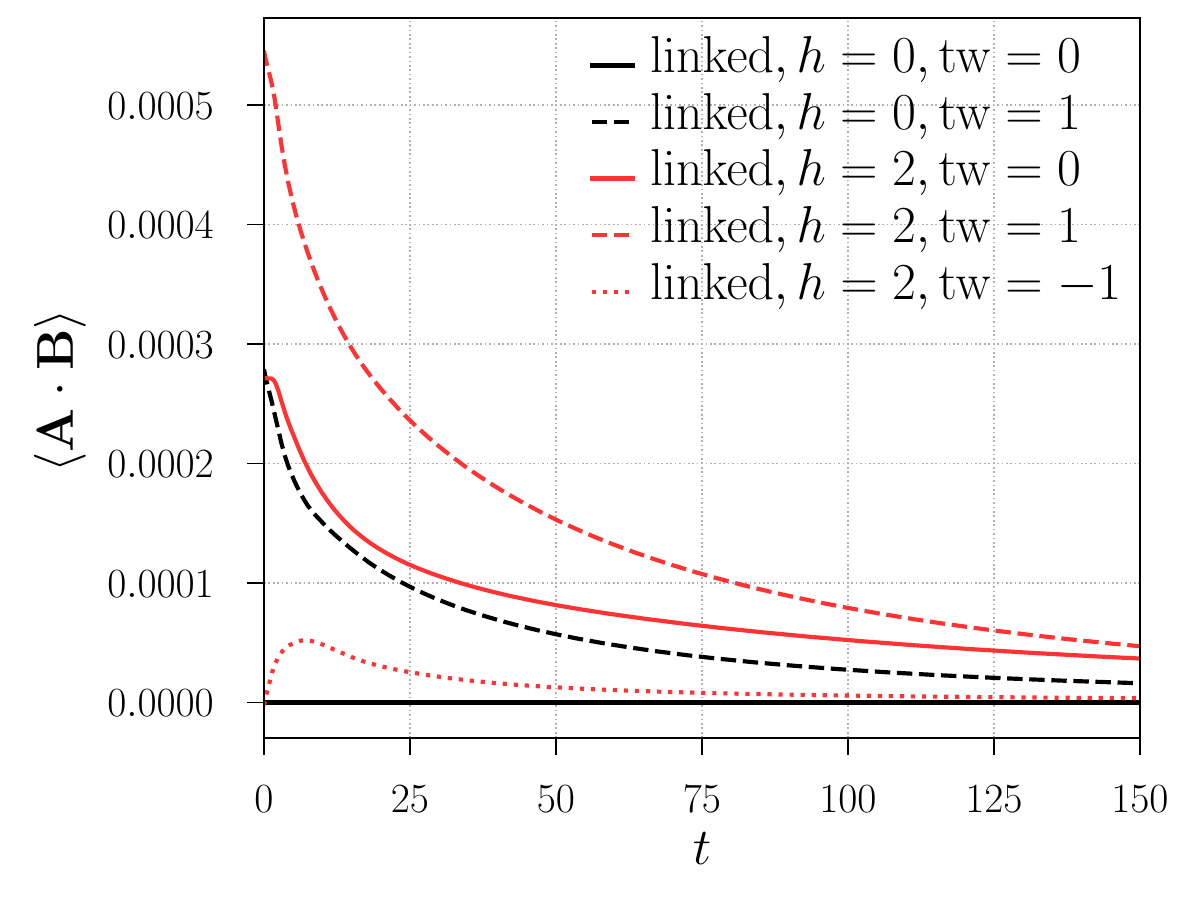}
\end{center}
\caption[]{
Mean magnetic helicity in dependence of time for the linked triple rings configurations using
different twist parameter ${\rm tw}$.
Cases labeled as $h = 0$ are the triple rings with no helicity if the twist
was zero.
Those labeled $h = 2$ are the ones with helicity if the twist was zero.
}\label{fig: abm_linked}
\end{figure}

While the twisted knots showed a significant change in magnetic helicity
at time scales much shorter than the diffusive time, the twisted
triple ring configurations exhibit a less dramatic behavior
(see \Fig{fig: abm_linked}).
This behavior is more reminiscent of what we would expect if we
assumed that helicity was conserved due to the small magnetic diffusivity $\eta$.
The reason is a much weaker alignment of the current density and the magnetic field,
even for the case of a strong internal twist.
\highlight{
Although currently we cannot give an explanation on why the triple rings
behave differently, it shows that an internal twist does not necessarily
lead to helicity decay or generation.
}

\subsubsection{Borromean Rings and $8_{18}$ Knot}

\begin{figure}[t!]\begin{center}
\includegraphics[width=\columnwidth]{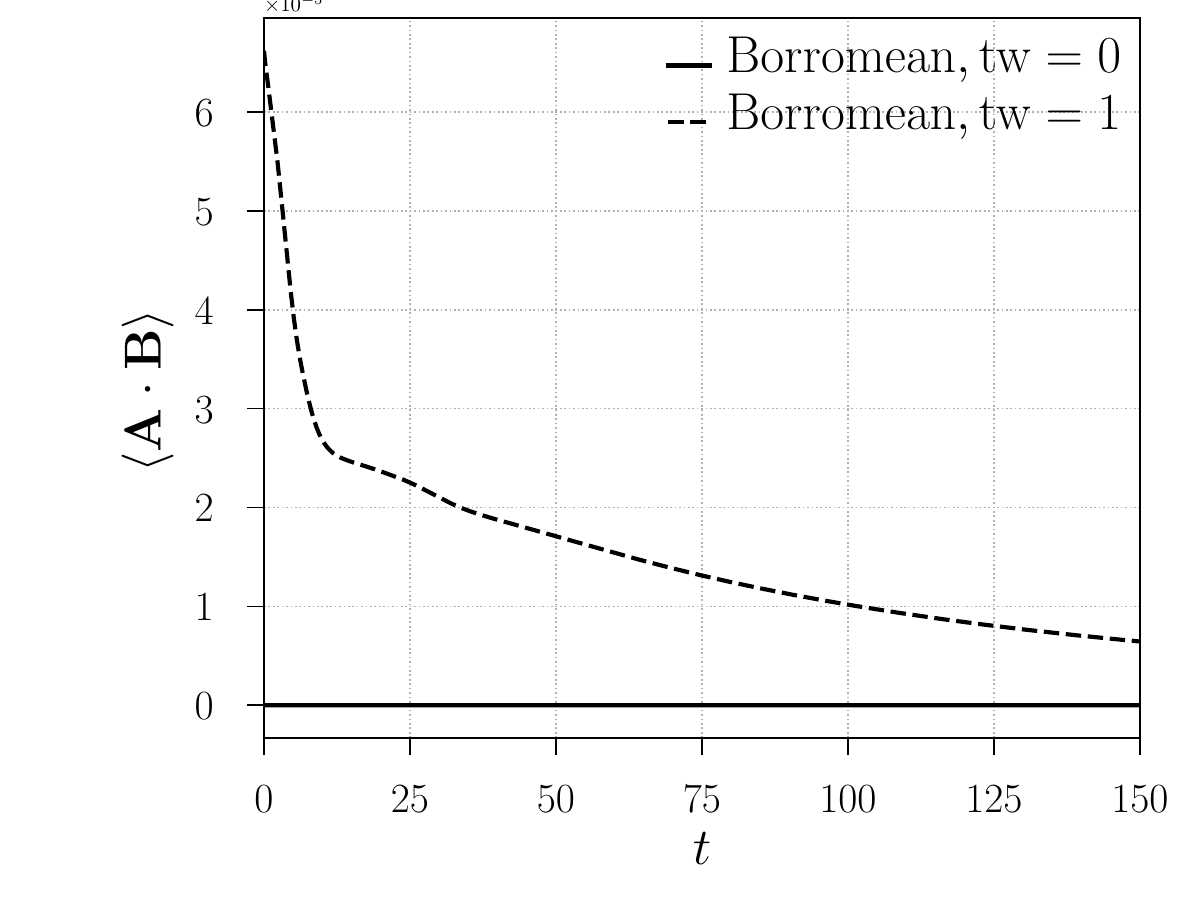} \\
\includegraphics[width=\columnwidth]{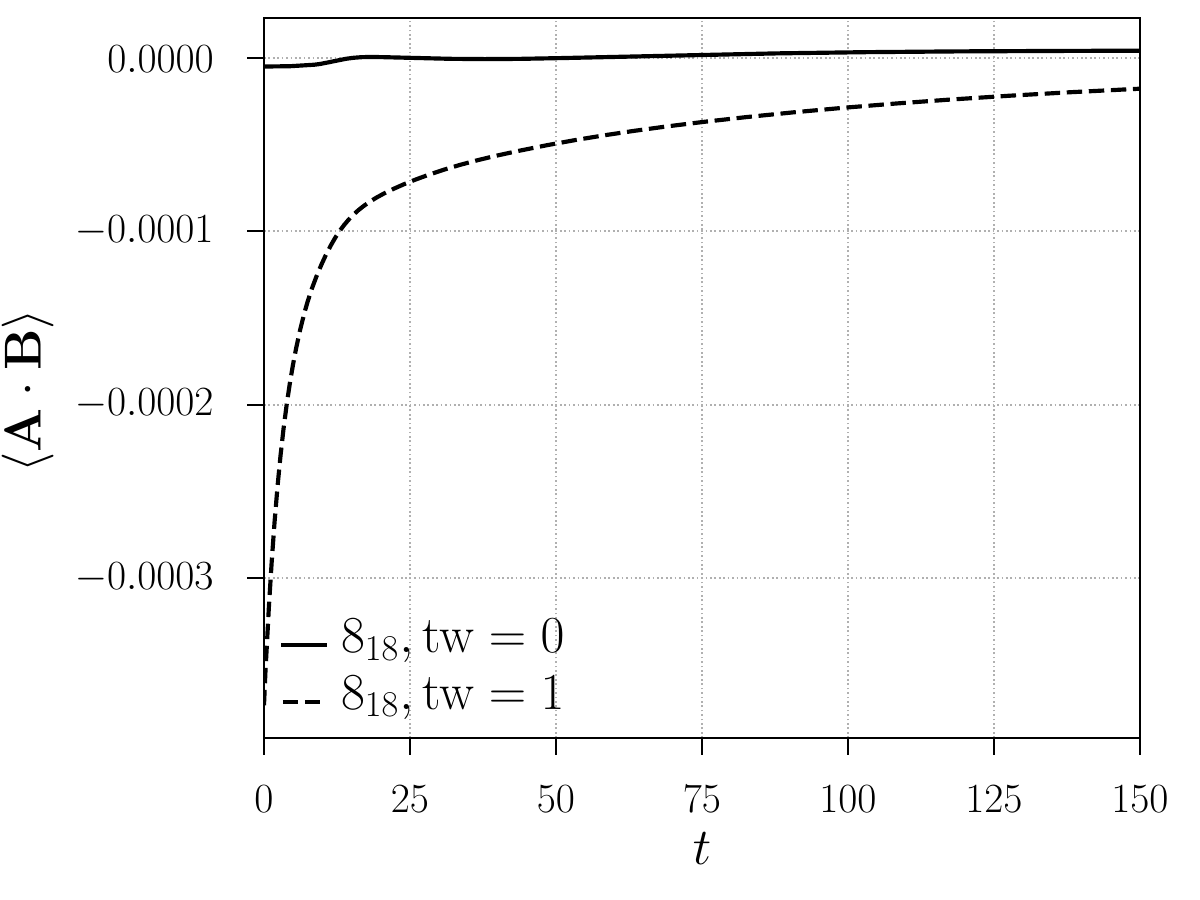}
\end{center}
\caption[]{
Mean magnetic helicity in dependence of time for the Borromean rings
(upper panel) and $8_{18}$ knot (lower panel) configurations using
different twist parameter ${\rm tw}$.
}\label{fig: abm_borromean_iucaa}
\end{figure}

The twisted Borromean rings and the $8_{18}$ knot both show similar
decay rates for the magnetic helicity (\Fig{fig: abm_borromean_iucaa}).
Within time scales as short as ca.\ $1\%$ of the diffusive time
scale the helicity already dissipates.
This should be contrasted to the near conservation
for the case of no internal twist.

\begin{figure}[t!]\begin{center}
\includegraphics[width=\columnwidth]{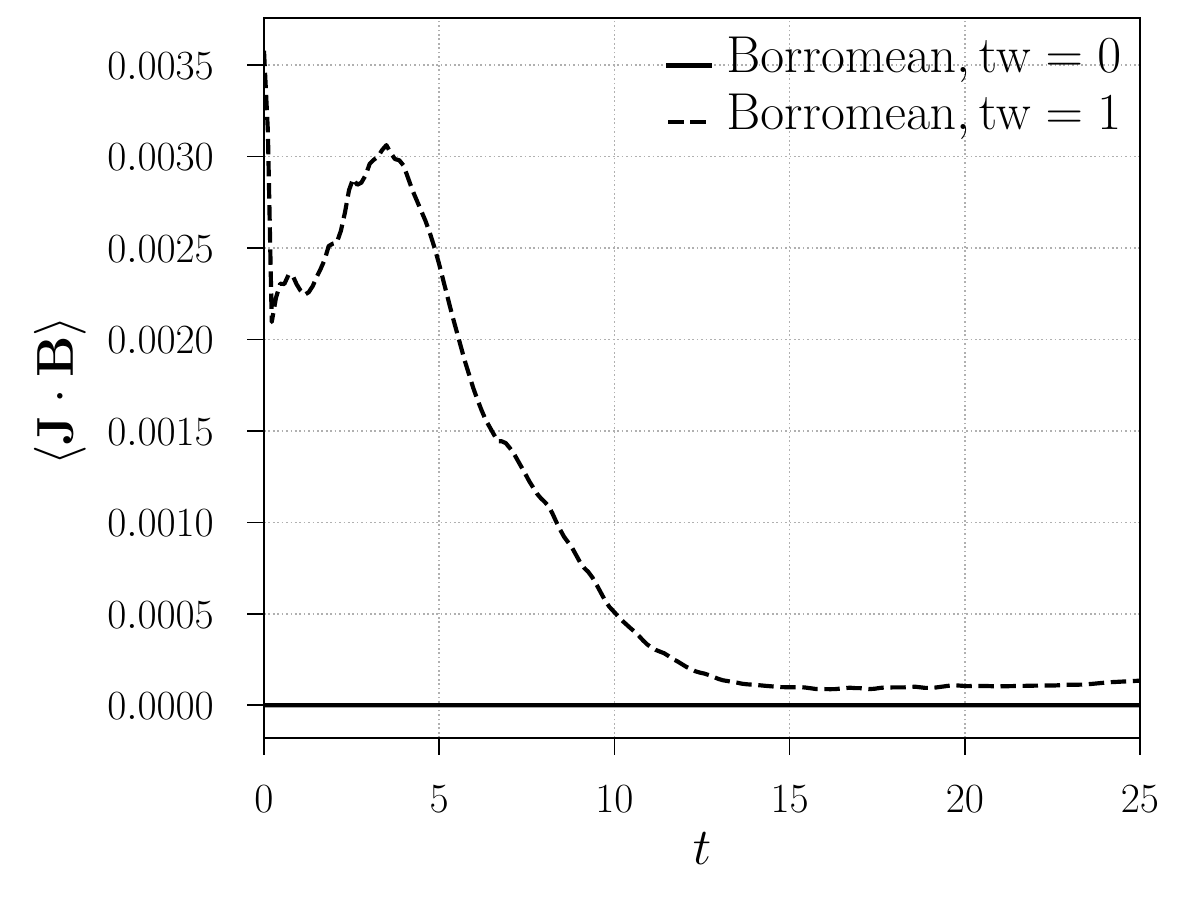} \\
\includegraphics[width=\columnwidth]{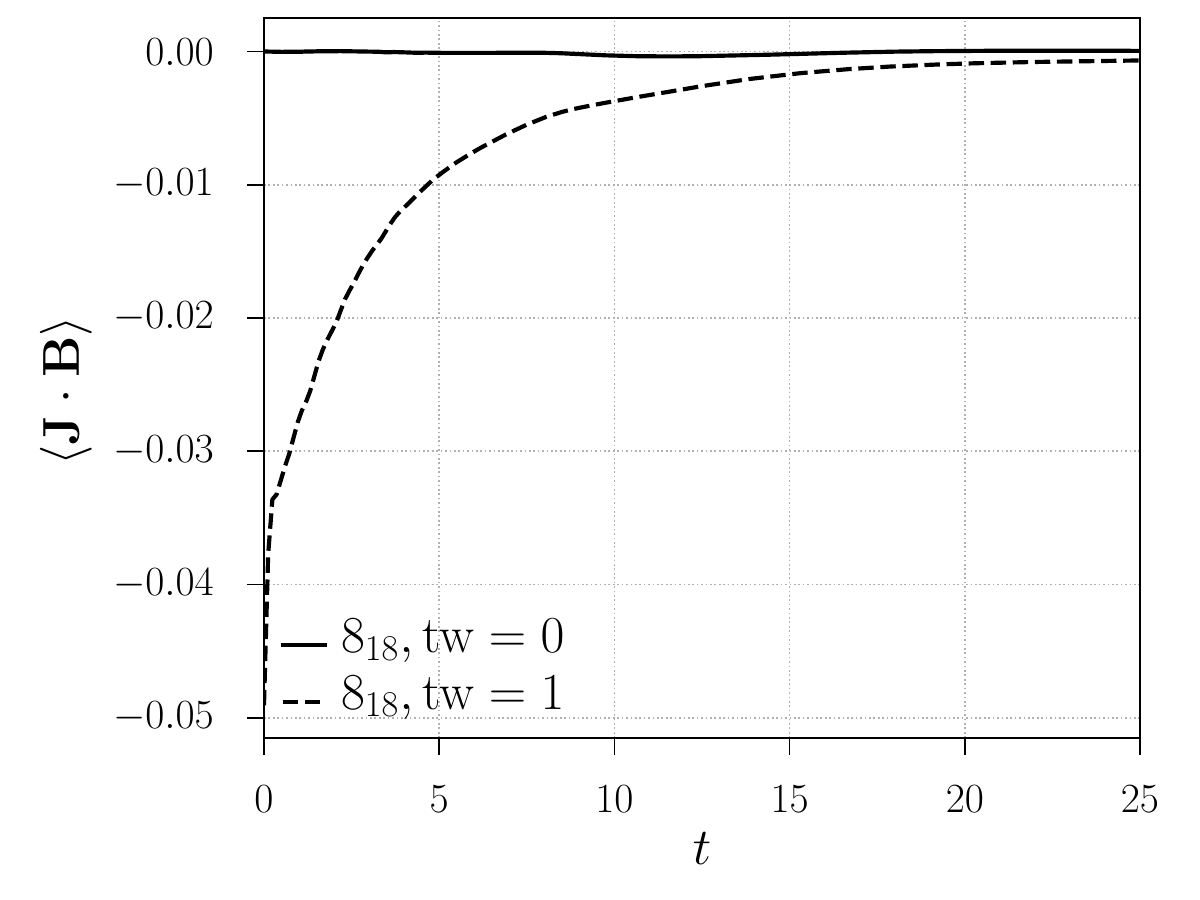}
\end{center}
\caption[]{
\highlight{
Mean current helicity in dependence of time for the Borromean rings
(upper panel) and $8_{18}$ knot (lower panel) configurations using
different twist parameter ${\rm tw}$.
}
}\label{fig: jbm_borromean_iucaa}
\end{figure}

\highlight{
The different behavior of the twist and no-twist case can be explained
by the alignment of $\JJ$ and $\BB$.
From \Fig{fig: jbm_borromean_iucaa} we see that the untwisted Borromean
rings and untwisted $8_{18}$ knot have a constantly low helicity production term.
This contrasts to the much larger values for the twisted cases.
So, similarly to the knots, the internal twist leads to helicity decay
due to a significant alignment of $\JJ$ and $\BB$.
}

\subsection{Magnetic Energy}

From previous studies \cite{fluxRings10, knotsDecay11} we know that magnetic
knots and links have a slow energy decay
compared to topologically trivial configurations.
It has been shown that, rather than the actual linking, the magnetic helicity content
is the determining factor for the speed of energy decay \cite{fluxRings10}.{

Here we have a set of test cases that can either corroborate those findings
or put a caveat to them.
If the magnetic helicity alone was the determining factor, then for the helical
cases we should see a slow decay
while for the twisted zero-helicity cases we should see a faster energy decay.

To test this hypothesis we plot the {\highlight{spatially averaged}
magnetic energy density for the triple ring
configurations with and without twist (\Fig{fig: b2m_linked})
and then normalize to its initial value ar $t = 0$.
\highlight{
This  spatial average is analogous to what we did for the helicities, i.e.
\EQ
\langle \BB^2 \rangle = \frac{1}{V} \int_{V} \BB^2 \ \dd V.
\EN
}
There we see that it is indeed the magnetic helicity that determines
the speed of the energy decay.
For the two non-helical configurations there is a steeper decay
than for the three helical cases.

\begin{figure}[t!]\begin{center}
\includegraphics[width=\columnwidth]{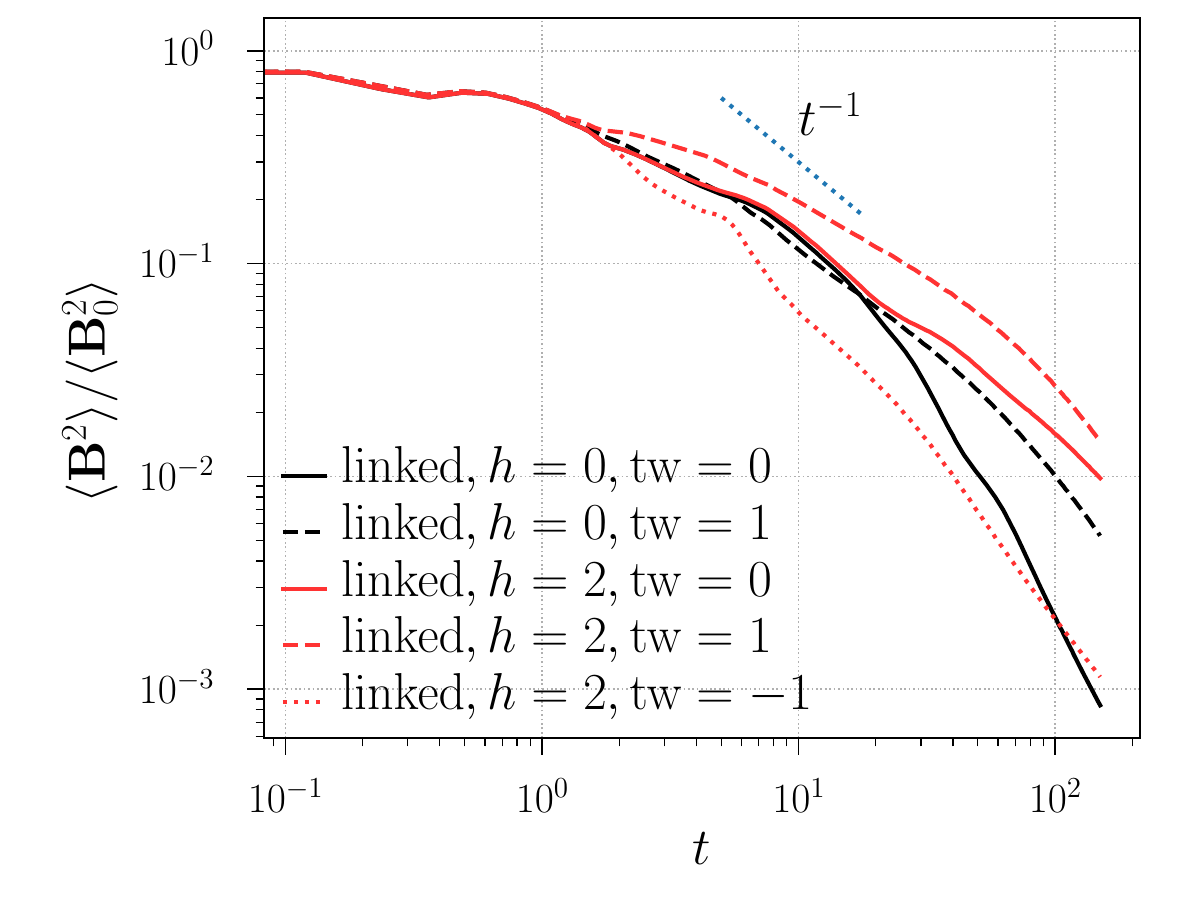}
\end{center}
\caption[]{
Normalized magnetic energy in dependence of time for the triple rings configurations using
different twist parameter ${\rm tw}$.
For guidance we add a power law of $t^{-1}$.
}\label{fig: b2m_linked}
\end{figure}

Of course we could have just picked the set of test cases that exactly proved
our hypothesis.
So, we also plot the magnetic energy for the knots (\Fig{fig: b2m_knots}).
There we can clearly see that the energy decay is of the same speed for all test
cases, independent of the type of knot (3, 4 or 5-foil) and the internal twist ${\rm tw}$.
Unlike the previous studies this suggests that it is the field line topology
manifested as knottedness and twist that is more of a determinant than
the {\em initial} helicity itself.
Here we emphasize {\em initial}, as the helicity can rapidly change even
in our low resistivity environment, as we have seen in the previous section.
This non-conservation is the reason for the similar energy decay.

\begin{figure}[t!]\begin{center}
\includegraphics[width=\columnwidth]{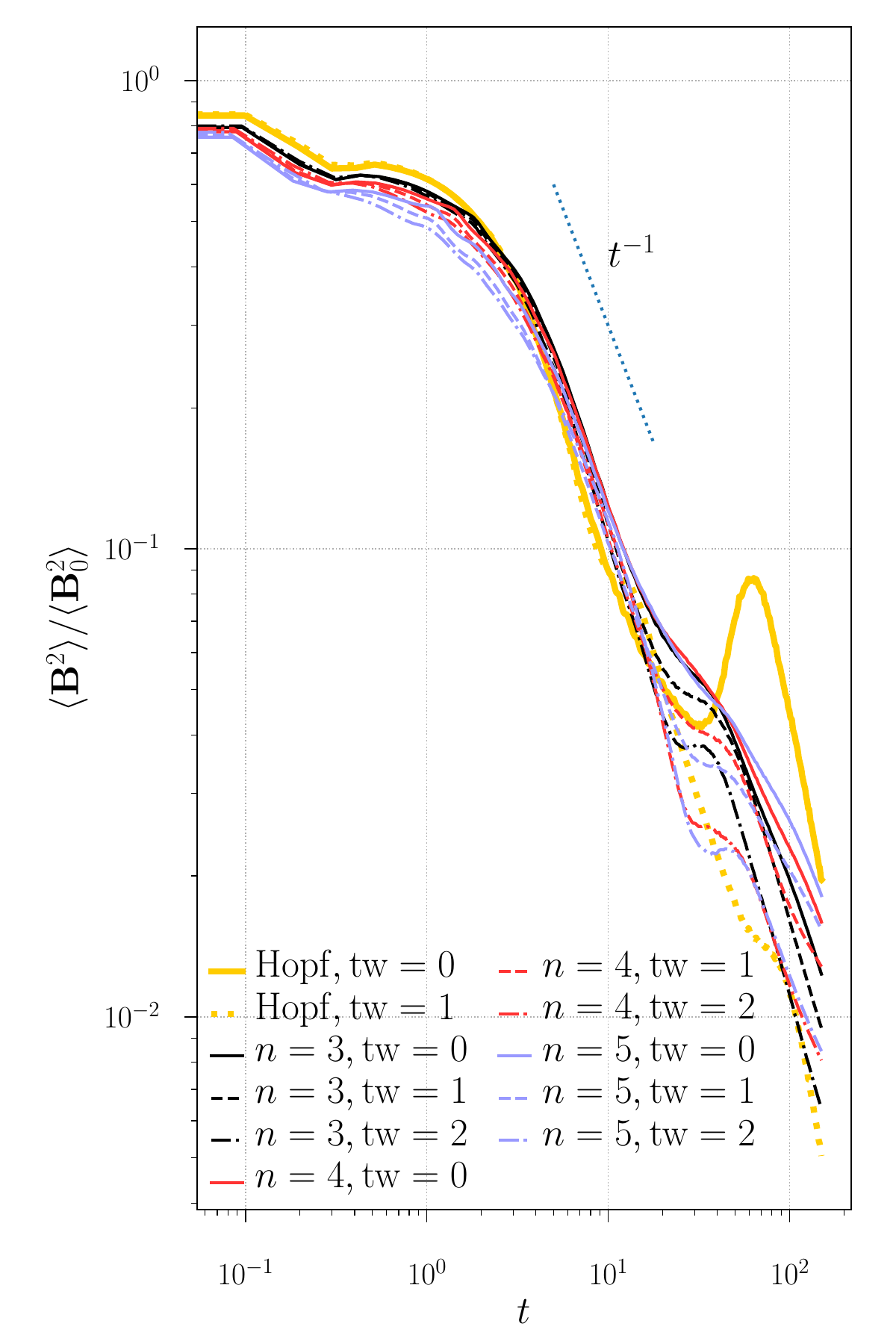}
\end{center}
\caption[]{
Normalized magnetic energy in dependence of time for the knot configurations using
different twist parameter ${\rm tw}$.
For guidance we add a power law of $t^{-1}$.
}\label{fig: b2m_knots}
\end{figure}

Since for the Borromean rings and the $8_{18}$ knot we also observe a rapid
loss of magnetic helicity we would expect that the helical and non-helical
cases exhibit a similar decay rate for the magnetic energy.
From \Fig{fig: b2m_borromean_iucaa} we see that this is indeed the case.
The decay rates are almost indistinguishable.

\begin{figure}[t!]\begin{center}
\includegraphics[width=\columnwidth]{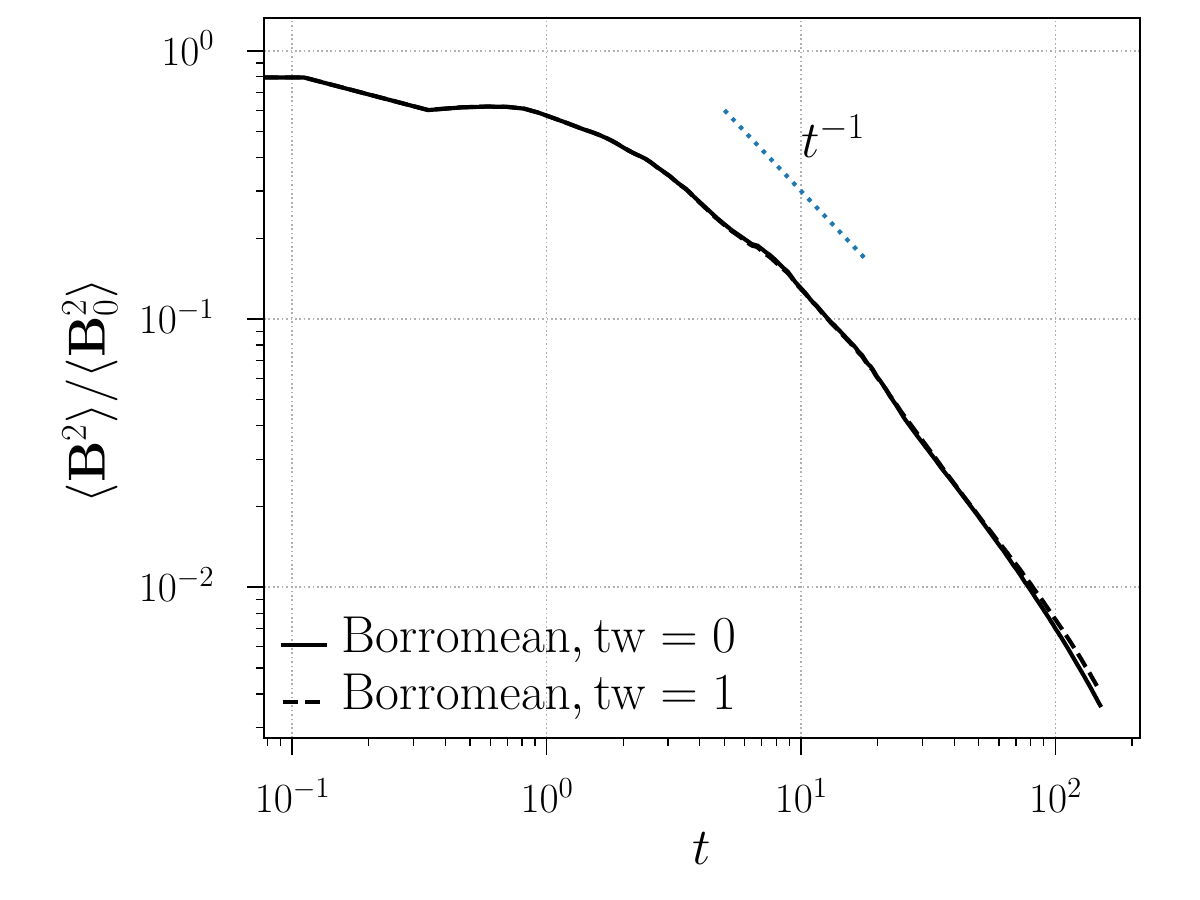} \\
\includegraphics[width=\columnwidth]{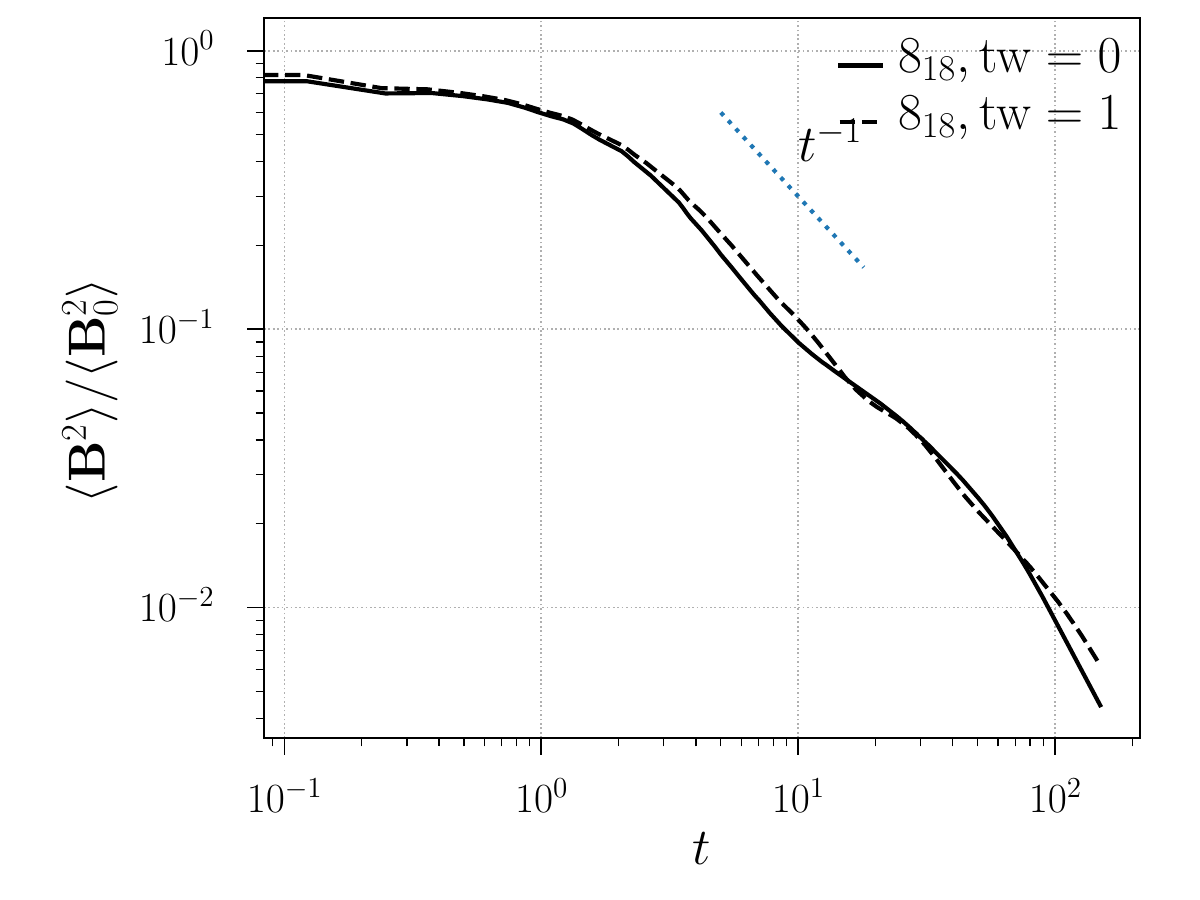}
\end{center}
\caption[]{
Normalized magnetic energy in dependence of time for the Borromean rings
configurations (top panel) and $8_{18}$ knot (bottom panel) using
different twist parameter ${\rm tw}$.
For guidance we add a power law of $t^{-1}$.
}\label{fig: b2m_borromean_iucaa}
\end{figure}

While the realizability condition provides a lower bound of the magnetic energy in presence
of magnetic helicity, an initially helical field is no guarantee that
the energy decays only slowly.
The observed helicity losses at time scales much shorter than the diffusive time
annul such decay restrictions.
We saw that only for the triple rings there is a relative conservation
of helicity and with that a slow energy decay.

\section{Effect of Low Magnetic Resistivity}

The $\JJ\cdot\BB$-term that is responsible for the generation and annihilation of magnetic
helicity, is clearly large enough to lead to significant changes in the helicity
within dynamical time scales.
Since we have a prefactor of the magnetic resistivity $\eta$ in equation
\eqref{eq: helicity_dissipation} we expect an insignificant change in helicity
in low resistivity systems, like the solar atmosphere or laboratory plasmas.

At the same time we know that with reduced resistivity, plasmas are more
turbulent and magnetic reconnection is more efficient.
For reconnection to happen we need alignment of $\JJ$ and $\BB$.
So, it is conceivable that in the low resistivity regime a reduced $\eta$ is compensated
by an increased $\JJ\cdot\BB$.

In this section we like to study the rate of change of magnetic helicity for low dissipation regimes.
We repeat the numerical experiment for the trefoil knot with strong internal twist,
i.e.\ with internal twist that changes the sign of the magnetic helicity.
We halve the magnetic resistivity $\eta$ to $5\times 10^{-4}$, as well as the viscosity $\nu$
to $5\times 10^{-4}$ in order to keep the Prandtl number at $1$.

We first observe that the alignment of the electric current density $\JJ$ and
the magnetic field $\BB$ is stronger with reduced magnetic diffusion
(see \Fig{fig: jbm_low_diff}).
At early times (until ca.\ $t = 5$) the difference is relatively small
by ca.\ $30\%$.
At later times, the relative difference is more of a factor of $4$,
where $\JJ\cdot\BB$ is small for both cases.
For the helicity production this quantity is relevant, but also carries
a prefactor of $\eta$, which is half for the low diffusion case.

\begin{figure}[t!]\begin{center}
\includegraphics[width=\columnwidth]{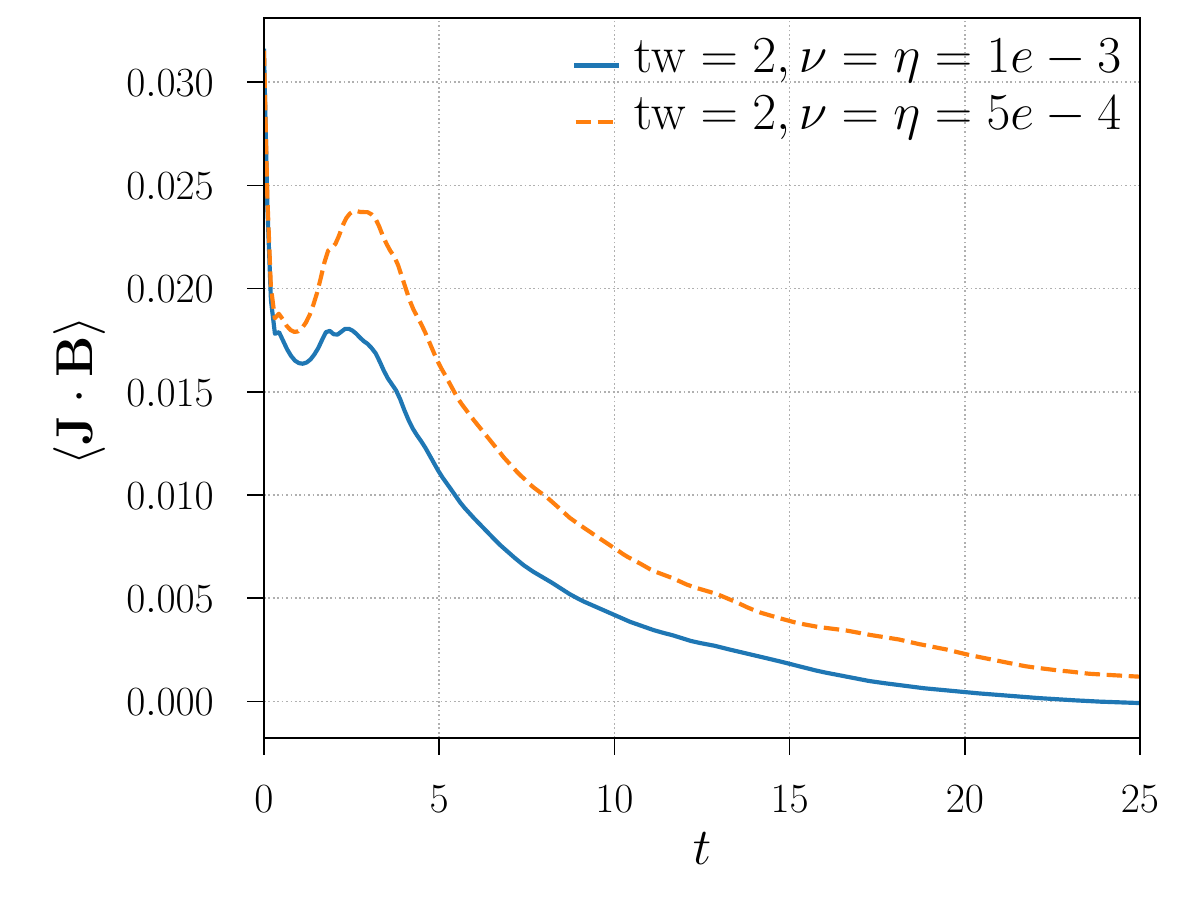}
\end{center}
\caption[]{
Alignment of the electric current density $\JJ$ and magnetic field
$\BB$ for the trefoil knot with twist $2$ and comparing high and
low diffusivity.
}\label{fig: jbm_low_diff}
\end{figure}

The observed increase of the $\JJ$--$\BB$ alignment is, however, not strong enough to compensate
for the halved prefactor $\eta$ in the helicity production.
For the helicity we observe a significantly smaller drop for the low diffusion case
(see \Fig{fig: abm_low_diff}).
From the realizability condition, we would therefore expect a smaller drop in
magnetic energy, at least for those times.
This is indeed what we observe (see \Fig{fig: b2m_low_diff}).
However, for later times there is a drop in magnetic helicity even for the
low diffusion case, which is why we also observe a significant drop in magnetic
energy at time $t > 40$.

\begin{figure}[t!]\begin{center}
\includegraphics[width=\columnwidth]{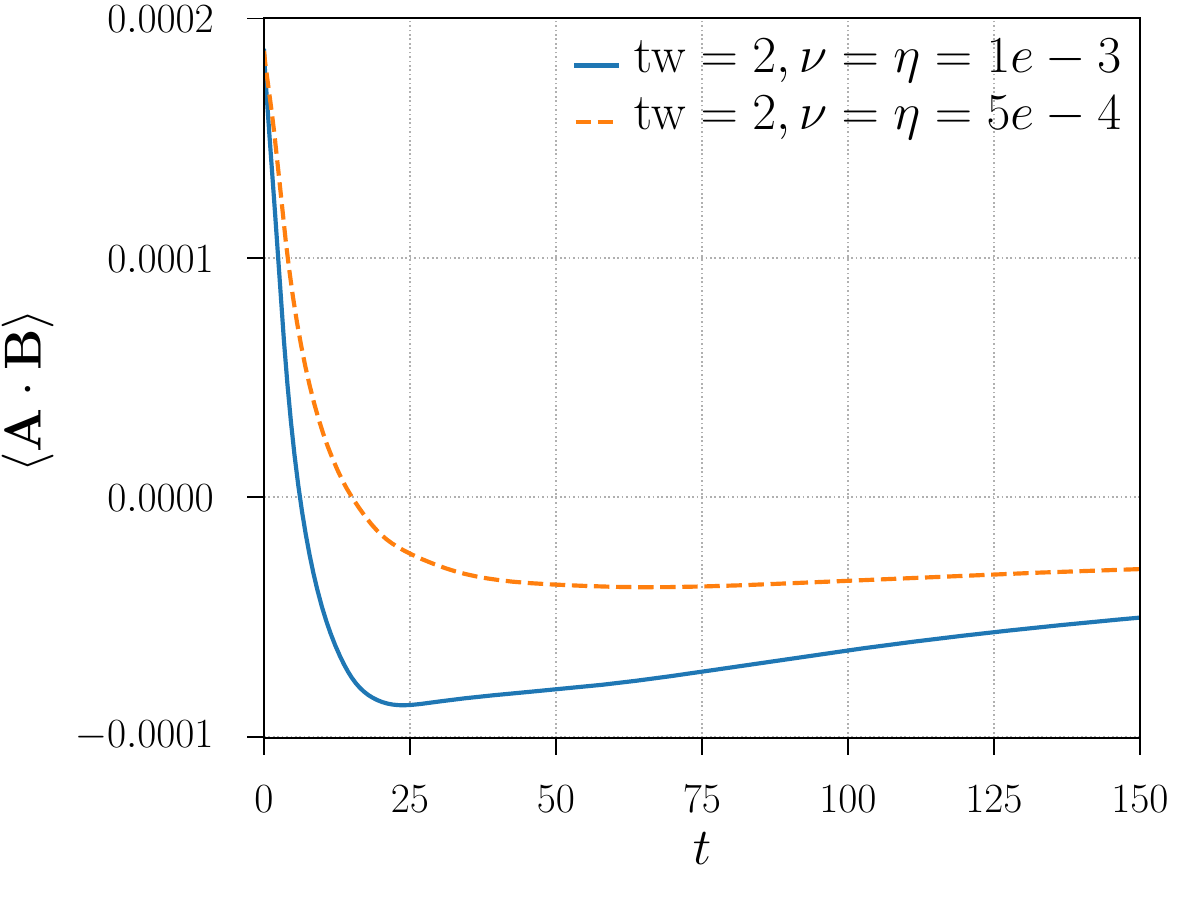}
\end{center}
\caption[]{
Mean magnetic helicity in dependence of time for the trefoil
knot with twist $2$ and comparing high and low diffusivity.
}\label{fig: abm_low_diff}
\end{figure}

\begin{figure}[t!]\begin{center}
\includegraphics[width=\columnwidth]{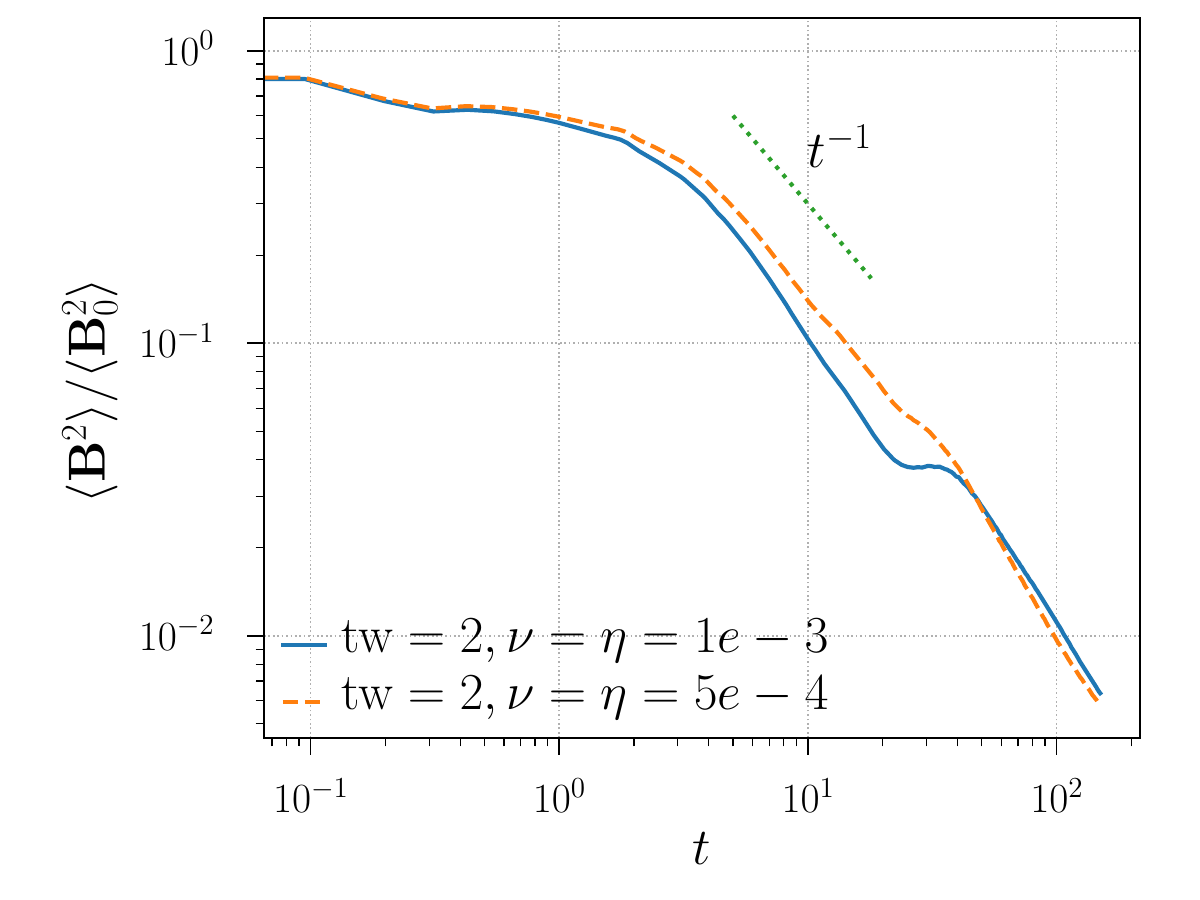}
\end{center}
\caption[]{
Normalized magnetic energy in dependence of time for the trefoil
knot with twist $2$ and comparing high and low diffusivity.
}\label{fig: b2m_low_diff}
\end{figure}

\section{Conclusions}

In our study we investigated the the robustness of the role of magnetic helicity
in the relaxation of topologically non-trivial magnetic flux tubes
in a plasma environment.
This was done by simulating the visco-resistive relaxation of
magnetic knots and links.
To contrast the effect of topology and magnetic helicity we vary
the internal twist of the fields.

While we confirm the importance of the realizability condition
as restriction of the field relaxation, the presence of magnetic helicity
does not automatically lead to a restricted decay.
One reason for this is the presence of source terms for the magnetic helicity.
Especially for fields with an internal twist we see that this term can
be so large to significantly change the helicity content within dynamical time scales.
With the loss of helicity, the restrictions on the relaxation
do not hold.

This has implications for solar magnetic fields, as they show
topologically non-trivial forms and can contain an internal twist.
Here we should keep in mind that through the twist the alignment of the
magnetic field with the electric current density is significant
and can potentially lead to changes in helicity.
However, to draw more precise conclusions for solar magnetic fields,
either in the convection zone, chromosphere or corona,
we would require a study that incorporates parameters found
in such environments, such as density and temperature gradients.

In order to apply our findings to laboratory or stellar plasmas we also
need to consider the scaling of all of our terms, particularly the
$\JJ$--$\BB$ alignment.
If we were to reduce the scale of our systems while keeping the magnetic
field strength $\BB$ constant, our currents $\JJ$  would increase due to
the larger spatial variations.
The same argument can be applied to the magnetic vector potential $\AAA$
which would decrease in strength.
The result is a smaller magnetic helicity density $h = \AAA\cdot\BB$ and a larger
$\JJ$--$\BB$ alignment.
The increased $\JJ$--$\BB$ alignment leads to an increased loss/generation rate of
magnetic helicity that is inversely proportional to the scaling.
So, a half size system would lose/generate the same amount of helicity in half the time.

However, if we ask about the loss of normalized (to time $0$) helicity we introduce
one more scaling by dividing by the initial magnetic helicity.
The loss of this quantity happens then on a quarter of the time
of the un-scaled system.
Furthermore, in turbulent scenarios it is usually the resistive time scale that is important,
which is $\tau_{\rm res} = L^2/\eta$.
With changing scale $L$ our normalized magnetic helicity would then change at
the same diffusive times as in the un-scaled system.

\begin{acknowledgments}
Celine Beck thanks the London Mathematical Society for supporting
her with one Undergraduate Bursary Award (grant number URB-2021-76).
This project has benefited from funding by the Deutsche Forschungsgemeinschaft
(DFG, German Research Foundation) through the research unit FOR 5409
"Structure-Preserving Numerical Methods for Bulk- and Interface Coupling of Heterogeneous Models (SNuBIC)"
(project number 463312734).
For the time dependent plots we made use of the Python plotting library
Matplotlib \cite{Hunter:2007}.
We thank the anonymous referees for their input, which helped improving this manuscript.
\end{acknowledgments}

\section*{Author Declarations}

\subsection*{Conflict of Interest}

The authors have no conflicts to disclose.

\subsection*{Author Contributions}

{\bf Simon Candelaresi:}
Conceptualization (lead);
Data curation (lead);
Formal analysis (lead);
Investigation (lead);
Methodology (lead);
Visualization (lead);
Writing – original draft (lead).
Software (supporting);
Validation (supporting);
{\bf Celine Beck:}
Software (lead);
Validation (lead);
Investigation (supporting);
Methodology (supporting);
Formal analysis (supporting);
Writing – original draft (supporting).

\section*{Data Availability Statement}

The data that support the findings of this study are available from the corresponding
author upon reasonable request.

\appendix

\section{Field Evolution}
\label{app: field_evolution}

For completeness we show the evolution of the knots that have been
mentioned only in passing.
Those are the $4$ and $5$-foil knots.
The qualitative evolution is the same as for the trefoil knot
(see \Fig{fig: streamlines_4foil} and \Fig{fig: streamlines_5foil})
with the appearance of a twisted structure even for the initially
non-helical case.

\begin{figure*}[t!]\begin{center}
\includefigure{n4_r256_tw0_t0}{0.65}{a) $t=0$, tw$=0$}
\includefigure{n4_r256_tw1_041975_t0}{0.65}{b) $t=0$, tw$=1$}
\includefigure{n4_r256_tw2_08283_t0}{0.65}{c) $t=0$, tw$=2$} \\
\includefigure{n4_r256_tw0_tf}{0.65}{d) $t=150$, tw$=0$}
\includefigure{n4_r256_tw1_041975_tf}{0.65}{e) $t=150$, tw$=1$}
\includefigure{n4_r256_tw2_08283_tf}{0.65}{f) $t=150$, tw$=2$}
\end{center}
\caption[]{
Initial magnetic streamlines for the $4$-foil knot configuration (top) using
different twist parameter ${\rm tw}$ such that the left has no twist,
the middle has a twist that reduces the helicity to zero and the
right a strong twist with opposite helicity to the left one.
The lower figures show the streamlines at time $150$.
}\label{fig: streamlines_4foil}
\end{figure*}

\begin{figure*}[t!]\begin{center}
\includefigure{n5_r256_tw0_t0}{0.65}{a) $t=0$, tw$=0$}
\includefigure{n5_r256_tw1_497275_t0}{0.65}{b) $t=0$, tw$=1$}
\includefigure{n5_r256_tw2_99621_t0}{0.65}{c) $t=0$, tw$=2$} \\
\includefigure{n5_r256_tw0_tf}{0.65}{d) $t=150$, tw$=0$}
\includefigure{n5_r256_tw1_497275_tf}{0.65}{e) $t=150$, tw$=1$}
\includefigure{n5_r256_tw2_99621_tf}{0.65}{f) $t=150$, tw$=2$}
\end{center}
\caption[]{
Initial magnetic streamlines for the $5$-foil knot configuration (top) using
different twist parameter ${\rm tw}$ such that the left has no twist,
the middle has a twist that reduces the helicity to zero and the
right a strong twist with opposite helicity to the left one.
The lower figures show the streamlines at time $150$.
}\label{fig: streamlines_5foil}
\end{figure*}

\nocite{*}
\bibliographystyle{plainurl}
\bibliography{references}

\end{document}